%% file: main.tex
\documentclass[12pt,draftclsnofoot,onecolumn]{IEEEtran}

\usepackage{algorithm,algorithmic,amsfonts,amsmath,amssymb,bbm,color,comment,epsfig,float,graphics,epstopdf,amsthm,mathtools,multirow,multicol,setspace,subcaption,tikz,url}

\usepackage[sort,noadjust]{cite}

\usepackage[acronym]{glossaries}

\usepackage{enumitem}
\setitemize{itemsep=0mm,topsep=0mm,parsep=0pt,partopsep=0pt}

\makeatletter
\newcommand\subparagraph{%
  \@startsection{subparagraph}{5}
  {\parindent}
  {3.25ex \@plus 1ex \@minus .2ex}
  {-1em}
  {\normalfont\normalsize\bfseries}}
\makeatother

\usepackage{titlesec}
\let\subparagraph\relax
\titlespacing{\section}{0pt}{5pt plus 2pt minus 1pt}{3pt plus 1pt minus 0pt}
\titlespacing{\subsection}{0pt}{4pt plus 2pt minus 1pt}{2pt plus 1pt minus 0pt}
\usepackage[font=small]{caption}

\newtheorem{theorem}{Theorem}
\newtheorem{lemma}{Lemma}

\newtheorem{proposition}{Proposition}

\newcommand{\rmP}{\mathrm{P}}
\newcommand{\rmR}{\mathrm{R}}

\newcommand{\Lc}{{\cal L}}

\newcommand{\ue}{\textnormal{\tiny{UE}}}
\newcommand{\nameP}{D2D-MAM}
\newcommand{\nameS}{D2D-SMAM}
\newcommand{\nameT}{D2D-TMAM}

\newacronym{ue}{UE}{user equipment}
\newacronym{bs}{BS}{base station}
\newacronym{d2d}{D2D}{device-to-device}
\newacronym{csit}{CSIT}{channel state information at the transmitter}
\newacronym{csi}{CSI}{channel state information}
\newacronym{pdf}{pdf}{probability density function}
\newacronym{mmwave}{mmWave}{millimeter wave}
\newacronym{los}{LoS}{line-of-sight}
\newacronym{nlos}{NLoS}{non-line-of-sight}
\newacronym{ula}{ULA}{uniform linear array}
\newacronym{fdd}{FDD}{frequency-division duplex}
\newacronym{awgn}{AWGN}{additive white Gaussian noise}
\newacronym{snr}{SNR}{signal-to-noise-ratio}
\newacronym{5g}{5G}{fifth-generation}

\input{./Definitions.tex}

\title{D2D-Aided Multi-Antenna Multicasting under Generalized CSIT}
\author{Placido Mursia,~\IEEEmembership{Student Member,~IEEE,}
    Italo Atzeni,~\IEEEmembership{Member,~IEEE}, \\ Mari Kobayashi,~\IEEEmembership{Senior Member,~IEEE,}
    and David Gesbert,~\IEEEmembership{Fellow,~IEEE}
    \thanks{P. Mursia and D. Gesbert are with the Communication Systems Department, EURECOM, France (email: \{placido.mursia, david.gesbert\}@eurecom.fr). I. Atzeni is with the Centre for Wireless Communications, University of Oulu, Finland (email: italo.atzeni@oulu.fi). M. Kobayashi is with the Technical University of Munich, Germany (email: mari.kobayashi@tum.de).}
    \thanks{The work of P.~Mursia was supported by Marie Sk\l{}odowska-Curie Actions (MSCA-ITN-ETN 722788 SPOTLIGHT). The work of I.~Atzeni was supported by the Marie Sk\l{}odowska-Curie Actions (MSCA-IF 897938 DELIGHT). The work of M.~Kobayashi and D.~Gesbert was supported by the French-German Academy towards Industry 4.0 (SeCIF project) under Institut Mines-Telecom. Part of this work has been presented at IEEE ICC 2019~\cite{Murs19} and at ASILOMAR 2019~\cite{Murs19_2}.}
    }

\begin{document}

\maketitle

\begin{abstract}
Multicasting, where a \gls{bs} wishes to convey the same message to several \glspl{ue}, represents a common yet highly challenging wireless scenario. In fact, guaranteeing decodability by the whole \gls{ue} population proves to be a major performance bottleneck since the \glspl{ue} in poor channel conditions ultimately determine the achievable rate. To overcome this issue, two-phase cooperative multicasting schemes, which use conventional multicasting in a first phase and leverage \gls{d2d} communications in a second phase to effectively spread the message, have been extensively studied. However, most works are limited either to the simple case of single-antenna \gls{bs} or to a specific \gls{csit} setup. This paper proposes a general two-phase framework that is applicable to the cases of perfect, statistical, and topological \gls{csit} in the presence of multiple antennas at the \gls{bs}. The proposed method exploits the precoding capabilities at the \gls{bs}, which enable targeting specific \glspl{ue} that can effectively serve as \gls{d2d} relays towards the remaining \glspl{ue}, and maximize the multicast rate under some outage constraint. Numerical results show that our schemes bring substantial gains over traditional single-phase multicasting and overcome the worst-\gls{ue} bottleneck behavior in all the considered \gls{csit} configurations.

\end{abstract}

\newpage
%=========================================================================
\section{Introduction}\label{sec:Intro}
\glsresetall
%=========================================================================

Multicast services, where a \gls{bs} needs to convey a common valuable message to a set of \glspl{ue}, arise naturally in many wireless scenarios \cite{Jin06,Sid06,Khi06,Sirk10,Hou09,Zho14}. Notable examples are wireless edge caching, where popular media are cached during off-peak hours and subsequently streamed via multicasting \cite{Mad14,Pas16}, and the broadcasting of mission-critical messages in vehicular networks \cite{Ara13}. However, it is well known that multicasting over wireless channels is hindered by the \textit{worst-user-kills-all} effect, whereby the multicast capacity vanishes as the number of \glspl{ue} $K$ increases for a fixed number of \gls{bs} antennas \cite{Jin06,Sid06}. In fact, since the message transmitted by the \gls{bs} must be decoded by all the \glspl{ue}, the multicast capacity is limited by the \glspl{ue} with the smallest fading gain and the latter tends to decrease with the system dimension. In particular, for the case of i.i.d. Rayleigh fading channels, the multicast capacity vanishes quickly as it scales inversely proportional to $K$ \cite{Jin06}. 

To overcome this issue, different approaches have been considered in the literature (e.g., \cite{Jin06,Ntr09,Meh13,Ngo18,Khi06,Sirk10,Hou09,Exp18,Zho14,Yan19,Yin14,Sant20}), which can be roughly classified into three groups. In the first group, a subset of \glspl{ue} in good channel conditions is selected to be served, whereas the \glspl{ue} in poor channel conditions are neglected \cite{Ntr09,Meh13}. However, not only does such an approach result in limited network coverage, but it also implies solving a combinatorial problem to find the best subset of \glspl{ue}. 
%A simpler yet suboptimal solution is to select \glspl{ue} with a thresholding scheme base on their fading gain. However, such approach results in limited performance. 
The second group exploits multiple antennas at the transmitter and the resulting channel hardening to mitigate the variance of the individual received signal power as the number of \glspl{ue} increases \cite{Jin06,Ngo18}. However, such an approach is based on the assumption of i.i.d. Rayleigh fading channels and requires that the number of \gls{bs} antennas grows at least as $\log(K)$. Lastly, the third group builds on the \gls{ue} cooperation enabled by \gls{d2d} links. Indeed, \gls{d2d} communications hold the potential to counteract the performance limitations of several emerging applications in \gls{5g} wireless systems such as multicasting, machine-to-machine communication, and cellular-offloading \cite{Asa14,Teh14,Boc14,Bas14,Gup15}. In the relevant case of multicasting, \gls{d2d} communications between the \glspl{ue} can be leveraged to overcome the vanishing behavior of the multicast capacity by dividing the total transmission time in two phases. Here, conventional multicasting occurs only in the first phase, where the \gls{bs} transmits at such a rate that the common message is received by a subset of \glspl{ue} in favorable channel conditions. Then, these \glspl{ue} act as opportunistic relays and cooperatively retransmit the message in the second phase. This approach has been extensively studied in the literature under specific \gls{csit} assumptions and by focusing on the simple case of single-antenna transmitter \cite{Khi06,Sirk10,Hou09,Exp18,Zho14,Yan19,Yin14,Sant20}, as detailed next.

%Theoretical analysis of two-phase cooperative multicasting can be found in \cite{Khi06,Sirk10,Hou09,Exp18}. More specifically, \cite{Khi06} shows that the Shannon capacity is non-vanishing in the case of dense network (i.e., a scenario in which the number of receivers increases over a fixed network area). In \cite{Sirk10}, the upper bound on the capacity is shown to grow with the number of receivers in the case of dense network while it is constant in the case of extended network (i.e., a scenario in which the number of receivers increases together with the network area). A similar analysis can be found in \cite{Hou09} for IEEE 802.16-based wireless metropolitan area networks. An analogous two-phase method is presented in \cite{Zho14}, which focuses on minimizing the total power consumption while guaranteeing a certain coverage under perfect \gls{csi} at the transmitter (\gls{csit}). Lastly, \cite{Sant20} proposes a two-phase scheme where the signal received in both phases is combined by exploiting only statistical \gls{csit}. As the number of \glspl{ue} grows, it is shown that a proper choice of the transmit powers can achieve a multicast rate that scales as $2 \log(K)$.

Theoretical analysis of two-phase cooperative multicasting can be found in \cite{Khi06,Sirk10,Hou09,Exp18,Sant20}. More specifically, \cite{Khi06} established the multicast capacity by using a two-phase cooperative scheme for a simple network with i.i.d. Rayleigh fading channels. The multicast scaling was analyzed in \cite{Sirk10} for two different network models, where the multicast capacity was shown to grow as $\log (\log (K))$ in the case of dense network (i.e., a scenario in which the number of receivers increases over a fixed network area) with spatially i.i.d. channels. Recently, \cite{Sant20} characterized the multicast scaling for a more general network topology (capturing the pathloss) and showed that, with statistical \gls{csit}, the average multicast rate increases as $\log (\log (K))$. A similar analysis can be found in \cite{Hou09} for IEEE~802.16-based wireless metropolitan area networks. Furthermore, \cite{Exp18} characterized the achievable multicast rate of an interactive scheme based on full-duplex and non-orthogonal cooperation links. Another two-phase scheme was presented in \cite{Zho14}, which focused on minimizing the total power consumption while guaranteeing a certain coverage under perfect \gls{csit}. On the other hand, \cite{Yan19} considered a two-layer multicast message structure with a high-priority, low-rate part and a low-priority, high-rate part, such that the \glspl{ue} who are able to decode the entire message assist the others by acting as opportunistic relays. The time allocation between the two phases was investigated in \cite{Yin14}, which showed that more time should be dedicated to the second phase as the \glspl{ue} move away from the \gls{bs}. Finally, a similar two-phase cooperative scheme with multiple antennas at the \gls{bs} was proposed in \cite{Mao20} in the context of broadcasting under perfect \gls{csit}. By exploiting rate splitting, this scheme forms a virtual common message to be multicast in the first phase and retransmitted via opportunistic relaying in the second phase.

In summary, existing works have demonstrated the benefits of two-phase cooperative schemes either for specific \gls{csit} configurations or for the simple case of single-antenna \gls{bs}. This motivates us to study the two-phase cooperative multicasting by exploiting multiple antennas at the \gls{bs} under various \gls{csit} configurations ranging from perfect \gls{csit} to topological \gls{csit}, where only the map of the network area and the \gls{ue} distribution are available at the \gls{bs}.

%All the works listed above consider two-phase cooperative multicasting in the simple case of single-antenna \gls{bs} and for specific \gls{csit} configurations. A similar two-phase cooperative scheme with multiple antennas at the \gls{bs} is proposed in \cite{Mao20} in the context of rate splitting and under the assumption of perfect \gls{csi} at both the transmitter and the receivers. Here, the message to be sent is split into common and private part, where the former is transmitted via multicasting from the \gls{bs} in the first phase and via opportunistic relaying in the second phase. 

%=========================================================================
\subsection{Contribution}
%=========================================================================

In this paper, we propose a general two-phase cooperative multicasting framework that leverages both multi-antenna transmission at the \gls{bs} and \gls{d2d} communications between the \glspl{ue}. In particular, we highlight how endowing the \gls{bs} with multiple antennas radically transforms the problem of cooperative multicasting. Indeed, the precoding capabilities at the \gls{bs} introduce additional degrees of freedom for spatial selectivity that, exploited together with the \gls{d2d} links, modify the nature and the performance of the two-phase schemes described in the previous section. However, this implies the joint optimization of the precoding strategy at the \gls{bs} and the multicast rate, which is, at first glance, highly complex to tackle: to the best of our knowledge, this is the first work that addresses such a scenario.

We consider a general system model (in terms of both channel model and network topology) and explicitly optimize the precoding strategy at the \gls{bs} and the multicast rate over the two phases. More specifically, we propose several schemes to tackle different \gls{csit} configurations, namely: \textit{i)}~perfect \gls{csit}, where the instantaneous channels are perfectly known; \textit{ii)}~statistical \gls{csit}, where only the long-term channel statistics are available; and \textit{iii)}~topological \gls{csit}, where only the map of the network area and the \gls{ue} distribution are accessible. Note that statistical \gls{csit} applies to scenarios with a large number of \glspl{ue} or limited feedback in frequency-division duplex mode, while topological \gls{csit} applies to scenarios where neither instantaneous nor statistical \gls{csit} is available and only the \gls{ue} distribution across the network can be considered for the optimization (see, e.g., \cite{Har09}). In addition, following \cite{Sant20}, we use the notion of target outage in the optimization of the multicast service, by which the multicast~rate is maximized while guaranteeing decodability by most \glspl{ue} up to the desired success level. In this way, we strategically avoid wasting resources on a small amount of \glspl{ue} with particularly unfavorable channel conditions \cite{Des18}. Numerical results show that the proposed schemes significantly outperform conventional single-phase multi-antenna multicasting in all the considered \gls{csit} configurations. Remarkably, they allow to effectively overcome the vanishing behavior of the multicast rate and achieve an increasing performance as the \gls{ue} population grows large.

The contributions of this paper are summarized as follows:
\begin{itemize}
    \item[$\bullet$] Assuming a general channel model and network topology, we propose a two-phase cooperative multicasting framework with multi-antenna transmission at the \gls{bs}. We tackle the joint optimization of the precoding strategy at the \gls{bs} and the multicast rate subject to some outage constraint. This framework is particularized to three different \gls{csit} configurations, i.e., perfect, statistical, and topological \gls{csit}. An interesting feature of our algorithms is to provide, as by-product, a selection of the \glspl{ue} that are best positioned to serve as \gls{d2d} relays to the remaining \glspl{ue} without the need for any explicit relay selection scheme.
    
    \item[$\bullet$] For the case of perfect \gls{csit}, we propose a low-complexity iterative algorithm that jointly selects a subset of \glspl{ue} to be served by the \gls{bs} in the first phase and optimizes the multicast rate while guaranteeing the desired success level. This algorithm, referred to as \textit{\nameP{}}, is shown to converge to a locally optimal solution. 
    
    \item[$\bullet$] For the case of statistical \gls{csit}, we propose a low-complexity algorithm that relies on long-term channel statistics without requiring costly instantaneous \gls{csit}, which is a major advantage in scenarios with a large number of \glspl{ue} or limited feedback. For this algorithm, referred to as \textit{\nameS{}}, we study the scaling of the resulting multicast rate as a function of the number of \glspl{ue} and \gls{bs} antennas and show that this is non-vanishing in the case of dense network.
    
    \item[$\bullet$] For the case of topological \gls{csit}, we propose an algorithm based on Monte Carlo sampling that relies uniquely on the map of the network area and the \gls{pdf} of the \gls{ue} locations. This approach is desirable in scenarios where neither instantaneous nor statistical \gls{csit} is available and only the \gls{ue} distribution across the network can be considered for the optimization. The proposed algorithm, referred to as \textit{\nameT{}}, runs the \nameP{} algorithm on several sets of \gls{ue} locations and channels generated according to the \gls{ue} distribution, and the outputs are averaged to obtain the actual precoding strategy at the \gls{bs} and multicast rate.
    
    \item[$\bullet$] We present a comprehensive numerical evaluation of the proposed schemes showing substantial gains compared to the reference single-phase multi-antenna multicasting in the three different \gls{csit} configurations.
\end{itemize}

%=========================================================================
\subsection{Outline and Notation}
%=========================================================================

The rest of the paper is organized as follows. Section~\ref{sec:Sysmodel} describes the system model. Section~\ref{sec:perfCSI} deals with the case of perfect \gls{csit} and introduces the \nameP{} algorithm. Section~\ref{sec:statCSI} tackles the case of statistical \gls{csit} and presents the \nameS{} algorithm. Section~\ref{sec:continuous} considers~the~case of topological \gls{csit} and proposes the \nameT{} algorithm. Then, Section~\ref{sec:NR} provides numerical results assessing the performance of the proposed schemes in the various \gls{csit} configurations. Finally, Section~\ref{sec:con} summarizes our contributions and draws some concluding remarks.

Throughout the paper, scalars are denoted by italic letters, while (column) vectors and matrices are denoted by boldface lowercase and uppercase letters, respectively. $\Compl$ represents the set of complex numbers, whereas $\Compl^{N \times M}$ denotes the set of $(N \times M)$-dimensional complex matrices. $(\cdot)^{\tran}$, $(\cdot)^{\herm}$, and $(\cdot)^*$ are the transpose, Hermitian transpose, and conjugate operators, respectively. $\1$ and $\0$ represent the all-one vector and the all-zero matrix, respectively, of proper dimensions. The $N$-dimensional identity matrix is denoted by $\I_N$, whereas $\e_n$ indicates its $n$th column. $\|\cdot\|$ represents the Euclidean norm for vectors, whereas $\Exp[ \, \cdot \, ]$ and and $\mathbbm{1}[ \, \cdot \, ]$ are the expectation operator and the indicator function, respectively. Furthermore, $[a_{1}, \ldots, a_{N}]$ denotes horizontal concatenation, whereas $\{a_{1}, \ldots, a_{N}\}$ or $\{ a_{n} \}_{n \in \setN}$ denote the set of elements in the argument. Lastly, $X \underset{}{\overset{\Pr}{\to}} \bar{X}$ denotes convergence in probability of the random variable $X$, whereas $f(\epsilon) \underset{\epsilon \to 0}{\sim} g(\epsilon)$ means that $\lim_{\epsilon \to 0} \frac{f(\epsilon)}{g(\epsilon)} = 1$.

%=========================================================================
\section{System Model} \label{sec:Sysmodel}
%=========================================================================

\begin{figure}[t!]
\centering
\includegraphics[scale=0.95]{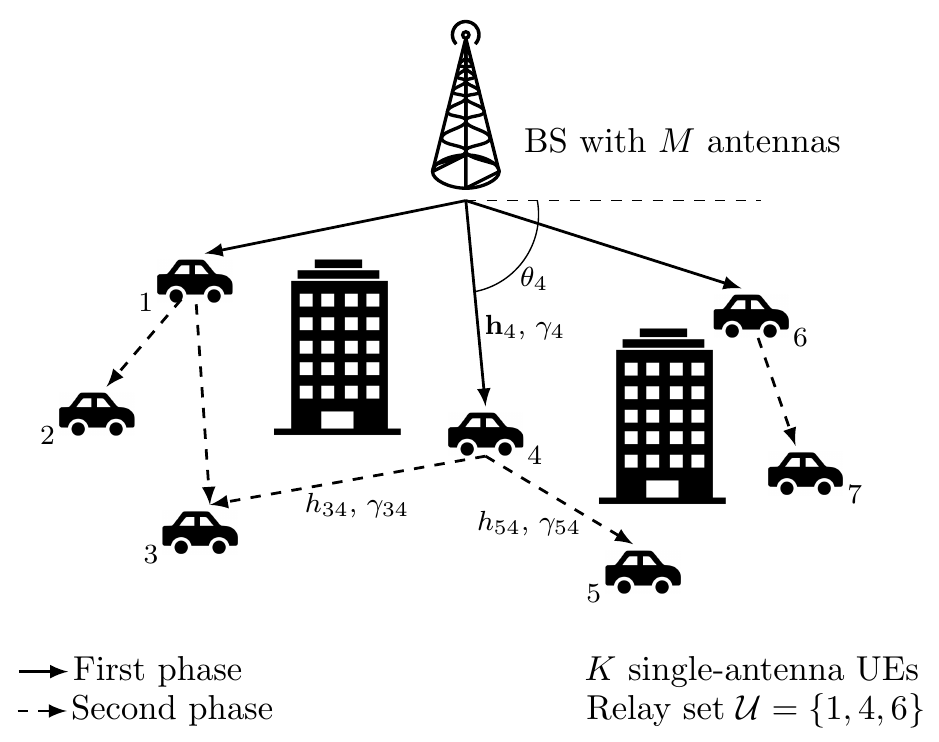}
\caption{A \gls{bs} equipped with $M$ antennas multicasts a common message to a subset of \glspl{ue} with a properly designed precoding strategy in the first phase (solid lines). The \glspl{ue} who successfully decode the message in the first phase retransmit it in the second phase to the remaining \glspl{ue} via \gls{d2d} links (dashed lines).}
\label{fig:sys_model}
\end{figure}

%=========================================================================
\subsection{Two-Phase Cooperative Multicasting} \label{subsec:two-phase}
%=========================================================================

We consider a wireless network where a \gls{bs} equipped with $M$ antennas aims at transmitting a common valuable message to a set of single-antenna \glspl{ue} $\setK \triangleq \{1,\ldots,K\}$, where $\h_k \in \Compl^{M\times 1}$ denotes the downlink channel between the \gls{bs} and \gls{ue}~$k$. The \glspl{ue} are also connected to each other via \gls{d2d} links in half-duplex mode, where $h_{jk}\in \Compl$ denotes the \gls{d2d} channel between \glspl{ue}~$j$ and $k$. We adopt a dense network scenario, i.e., where the number of receivers increases over a fixed network area, and we assume that $K \gg M$. For the sake of simplicity, we follow \cite{Khi06,Sirk10} and focus on a cooperative scheme divided into two phases of equal length. Such a scheme is depicted in Fig.~\ref{fig:sys_model} and the two phases are described next.
\begin{itemize}
    \item[\textit{1)}] \textbf{First phase.} The \gls{bs} transmits the message $\x\in \Compl^{M\times 1}$ at rate $r$, referred to as \textit{multicast rate}, and with transmit covariance $\Gammab \triangleq \Exp[\x\x^{\herm}]$, with $\tr(\Gammab) \leq 1$. The receive signal at \gls{ue}~$k$ in the first phase is given by
    \begin{align} \label{eq:yk1}
        y_{k,1} \triangleq \sqrt{\xi_0} \h_k^{\herm} \x + n_k \in \Compl
    \end{align}
    where $\xi_0$ is the transmit power at the \gls{bs} and, since we assume the \gls{awgn} noise term $n_k$ to be distributed as $\mathcal{CN}(0,1)$, it can be interpreted as the transmit \gls{snr} at the \gls{bs}. The message is decoded by \gls{ue}~$k$ if its achievable rate in the first phase is greater than or equal to the multicast rate $r$, i.e., if $\log_2(1+\xi_0 \h_k^{\herm}\Gammab\h_k) \geq r$. We define the subset of \glspl{ue} whose achievable rate in the first phase is at least $r$ for a given transmit~covariance~as
    \begin{align}
        \setU \triangleq \big\{ k\in \setK : \log_2(1 + \xi_0 \h_k^{\herm}\Gammab\h_k) \geq r \big\}.
    \end{align}
    
    \item[\textit{2)}] \textbf{Second phase.} The \glspl{ue} who were able to decode the message in the first phase jointly retransmit the message in an isotropic fashion, thus acting as opportunistic relays.\footnote{We assume that the \glspl{ue} retransmit the message with fixed power and do not perform any power control in the second phase.} Hence, the receive signal at \gls{ue}~$k$ in the second phase is a non-coherent sum of the \gls{d2d} transmit signals and is given by
    \begin{align}
        y_{k,2} = \sum_{j \in \setU} \sqrt{\xi_j} h_{jk} x_j + n_k \in \Compl, \quad \forall k\in \setK\setminus\setU
    \end{align}
    where $\xi_j$ is the transmit power at \gls{ue}~$j$ and can be interpreted as the transmit \gls{snr} at \gls{ue}~$j$ (cf. \eqref{eq:yk1}); moreover, $x_j$ is the message transmitted by \gls{ue}~$j$, with $\Exp[|x_j|^2] =1$. The message is successfully decoded by \gls{ue}~$k$ if its achievable rate in the second phase is greater than or equal to $r$, i.e., if $\log_2 \big( 1+|\sum_{j \in \setU} \sqrt{\xi_j} h_{jk}|^2 \big) \geq r$.
\end{itemize}

%=========================================================================
\subsection{Single-Phase Multicasting} \label{subsec:baseline}
%=========================================================================

As a special case of the above, we describe a single-phase multicasting scheme, which we refer to as \textit{baseline scheme}. This will serve as a means to assess the benefits brought by adding a second phase of \gls{d2d} communications to traditional multi-antenna multicasting. In this scheme, the \gls{bs} simply transmits the common message aiming at reaching all the \glspl{ue}. The receive signal at \gls{ue}~$k$ is the same as \eqref{eq:yk1} and the multicast capacity is given by (see \cite{Jin06})
\begin{align}
    C(\H) & \triangleq \max_{\Gammab \succeq \0 \; : \; \tr(\Gammab)\leq 1} \min_{k\in \setK} \log_2(1 + \xi_0 \h_k^{\herm}\Gammab\h_k) \label{eq:P_multicast} \\
    & = \log_2 \Big(1 + \xi_0 \max_{\Gammab \succeq \0 \; : \; \tr(\Gammab)\leq 1} \min_{k\in \setK} \h_k^{\herm}\Gammab\h_k\Big) \label{eq:P_multicast1}
\end{align}
where $\H = [\h_1,\ldots,\h_K]\in \Compl^{M\times K}$. Although a closed-form expression of the multicast capacity is not available, $C(\H)$ is convex in $\Gammab$ and, therefore, it can be computed via semidefinite programming. The main drawback of this single-phase scheme is that the multicast capacity is limited by the \gls{ue} with the worst channel conditions. In particular, for the case of i.i.d. Rayleigh fading channels and when the number of \gls{bs} antennas $M$ is fixed, the multicast capacity scales as $K^{-1/M}$ \cite{Jin06}.

%=========================================================================
\subsection{Channel Model} \label{subsec:CM}
%=========================================================================

Following the \gls{mmwave} one-ring channel model (see, e.g., \cite{Uwa20} and references therein), let us express the direct channel to \gls{ue}~$k$ as
\begin{align} \label{eq:channel}
    \h_{k} \triangleq \eta_{k} \sqrt{\gamma_{k}} \a_k \in \Compl^{M\times 1}
\end{align}
where $\eta_{k} \sim \setC \setN (0,1)$ is the small-scale fading coefficient, $\gamma_{k}$ is the average channel power gain, and $\a_k \in \Compl^{M \times 1}$ is the array response vector at the \gls{bs} for the steering angle $\theta_k$, with $||\a_k||^2 =M$. Here, we have $\gamma_k= d_k^{-\alpha}$ in case of \gls{los} conditions and $\gamma_k = d_k^{-\beta}$ in case of \gls{nlos} conditions, where $d_k$ denotes the distance between the \gls{bs} and \gls{ue}~$k$ and $\alpha$ (resp. $\beta$) is the \gls{los} (resp. \gls{nlos}) pathloss exponent. For simplicity, we assume that the \gls{bs} is equipped with a \gls{ula}, such that
\begin{align}
    \a_k = [1, e^{-j2\pi\delta\cos(\theta_k)}, \ldots, e^{-j2\pi\delta(M-1)\cos(\theta_k)}]^{\tran} \in \Compl^{M\times 1} \label{eq:ULA}
\end{align}
where $\delta = 0.5$ is the ratio between the antenna spacing and the signal wavelength. On the other hand, the \gls{d2d} channel between \glspl{ue}~$k$ and $j$ is represented as
\begin{align} \label{eq:channel_d2d}
    h_{jk} \triangleq \eta_{jk} \sqrt{\gamma_{jk}} \in \Compl
\end{align}
where $\eta_{kj} \sim \mathcal{CN}(0,1)$ is the small-scale fading coefficient and $\gamma_{jk}$ is the average channel power gain. Here, we have $\gamma_{jk} = d_{kj}^{-\alpha}$ in case of \gls{los} conditions and $\gamma_{jk} = d_{kj}^{-\beta}$ in case of \gls{nlos} conditions, where $d_{jk}$ denotes the distance between \glspl{ue}~$k$ and $j$ (cf. \eqref{eq:channel}).

%=========================================================================
\subsection{\gls{csit} Configurations} \label{subsec:CSIT}
%=========================================================================

In this paper, we consider several configurations of \gls{csit} that may be available at the \gls{bs} under different application scenarios.
\begin{itemize}
    \item[\textit{i)}] \textbf{Perfect \gls{csit} [Section~\ref{sec:perfCSI}].} The knowledge of both the direct channels, i.e., $\{\h_k\}_{k \in \setK}$, and the \gls{d2d} channels, i.e., $\{h_{jk}\}_{k,j \in \setK}$, is assumed.
    
    \item[\textit{ii)}] \textbf{Statistical \gls{csit} [Section~\ref{sec:statCSI}].} The knowledge of the \gls{ue} locations is assumed. From this information, the \gls{bs} can extract long-term statistics such as the average channel power gains of both the direct channels, i.e., $\{\gamma_k\}_{k \in \setK}$, and the \gls{d2d} channels, i.e., $\{\gamma_{jk}\}_{k,j \in \setK}$, together with the steering angles $\{\theta_k\}_{k \in \setK}$.
    
    \item[\textit{iii)}] \textbf{Topological \gls{csit} [Section~\ref{sec:continuous}].} The knowledge of the map of the network area, i.e., the location and size of the obstacles (such as buildings) within its coverage area, and of the \gls{pdf} of the \gls{ue} locations is assumed.
\end{itemize}
The above configurations correspond to settings with decreasing requirements on the information available at the \gls{bs}. While configuration~i) is relevant for the case of moderate (or finite) number of \glspl{ue} and low mobility, configuration~iii) is relevant for the case of large number of \gls{ue} and high mobility: for instance, these features arise in vehicular networks, where the \gls{bs} multi-antenna beam pattern ought to be designed on the basis of a city map and road traffic distribution. Lastly, configuration~ii) can be considered as an intermediate case between i) and iii).

%=========================================================================
\subsection{Performance Metrics} \label{subsec:metrics}
%=========================================================================

We propose two different performance metrics in terms of service reliability. In order to reflect the inherent difficulty to \emph{guarantee} a given data rate in a wireless setting with uncertainties on the channel conditions across the \glspl{ue}, we introduce the target outage $\epsilon \in [0,1)$, which describes the trade-off between the multicast rate and the reliability level at which we can maintain such a rate. Furthermore, let $\rmP_{k,1}(r, \Gammab)$ and $\rmP_{k,2}(r, \Gammab)$ denote the probabilities that \gls{ue}~$k$ successfully decodes in the first and in the second phase, respectively.
\begin{itemize}
    \item[\textit{a)}] \textbf{Average multicast rate.} We define the \textit{average success probability} as the probability that a randomly chosen \gls{ue} successfully decodes over the two phases, which is given by
    \begin{align} \label{eq:av_P}
        \rmP_{\mathrm{A}}(r, \Gammab) \triangleq \frac{1}{K} \sum_{k \in \setK}\big[\rmP_{k,1}(r, \Gammab) + \big( 1-\rmP_{k,1}(r, \Gammab) \big) \rmP_{k,2}(r, \Gammab)\big].
    \end{align}
    Hence, the \textit{average multicast rate} is defined as the maximum transmission rate at which a randomly chosen \gls{ue} successfully decodes with probability at least $1-\epsilon$ over the two phases, which can be expressed as
    \begin{align} \label{eq:R_av}
        \rmR_{\mathrm{A}}(r,\Gammab) \triangleq \frac{1}{2} r \quad \textrm{with}~r~\textrm{solution to}~\rmP_{\mathrm{A}}(r,\Gammab)\geq 1-\epsilon.
    \end{align}

    \item[\textit{b)}] \textbf{Outage multicast rate.} Let us introduce the binary variables $z_{k,1}(r, \Gammab)$ and $z_{k,2}(r, \Gammab)$, which are equal to $1$ if \gls{ue}~$k$ successfully decodes in the first and in the second phase, respectively, and to $0$ otherwise. Furthermore, let $\z_1(r,\Gammab) \triangleq [z_{1,1}(r,\Gammab)\ldots z_{K,1}(r,\Gammab)]$. We define the \textit{joint success probability} as the probability that all the \glspl{ue} successfully decode over the two phases, which is given by
    \begin{align} \label{eq:Prob_ph2J}
        \rmP_{\mathrm{J}}(r,\Gammab) \triangleq \Exp \bigg[ \prod_{k\in\setK} \Pr \bigg[ \log_2\bigg(1+\bigg\rvert \sum_{j\neq k} \xi_j h_{kj} \bigg\rvert^2 \bigg) \geq r \big( 1 - z_{k,1}(r,\Gammab) \big) \bigg\rvert \z_1(r,\Gammab) \bigg] \bigg]. 
    \end{align}
    Hence, the \textit{outage multicast rate} is defined as the maximum transmission rate at which all the \glspl{ue} successfully decode with probability at least $1-\epsilon$ over the two phases, which can be expressed as
    \begin{align} \label{eq:R_out}
        \rmR_{\mathrm{O}}(r,\Gammab) \triangleq \frac{1}{2} r \quad \textrm{with}~r~ \textrm{solution to}~\rmP_{\mathrm{J}}(r,\Gammab) \geq 1-\epsilon.
    \end{align}
\end{itemize}

%=========================================================================
\subsection{Problem Formulation} \label{subsec:Prob}
%=========================================================================

Our objective is to jointly optimize the multicast rate $r$ and the transmit covariance $\Gammab$ under one of the above outage constraints over the two phases. Such a problem can be formalized as\footnote{The factor $\frac{1}{2}$ in the objective describes the equal time division between the two phases and is irrelevant for the optimization.}
\begin{align}\label{eq:Problem}
\begin{array}{cl}
     \displaystyle \max_{r>0, \Gammab\succeq \0} & \displaystyle \frac{1}{2} r\\
     \mathrm{s.t.} & \tr(\Gammab)\leq 1,\\
     & \rmP_{\mathrm{T}}(r,\Gammab) \geq 1-\epsilon
\end{array}
\end{align}
where $\mathrm{T}\in \{\mathrm{A}, \mathrm{J}\}$. Hence, when $\mathrm{T} = \mathrm{A}$, we recover the average multicast rate $\rmR_{\mathrm{A}}(r,\Gammab)$ defined in \eqref{eq:R_av} and, when $\mathrm{T} = \mathrm{J}$, we recover the outage multicast rate $\rmR_{\mathrm{O}}(r,\Gammab)$ defined in \eqref{eq:R_out}. Note that problem \eqref{eq:Problem} is non-convex in both optimization variables due to the non-convex outage constraint and is thus highly complex to solve. In the following, we detail our proposed methods to tackle problem~\eqref{eq:Problem} in the three \gls{csit} configurations described in Section~\ref{subsec:CSIT}.

%=========================================================================
\section{\gls{d2d}-Aided Multi-Antenna Multicasting with Perfect \gls{csit}} \label{sec:perfCSI}
%=========================================================================

In this section, we consider the case where all the direct channels, i.e., $\{\h_k\}_{k \in \setK}$, and all the \gls{d2d} channels, i.e., $\{h_{jk}\}_{k,j \in \setK}$, are perfectly known at the \gls{bs}. For each \gls{ue}~$k$, let us define the binary variables
\begin{align}
    \label{eq:Zk1} z_{k,1}(r, \Gammab) & \triangleq \mathbbm{1} \big[ \log_{2} (1 + \xi_0 \h_{k}^{\herm} \Gammab \h_{k}) \geq r \big], \\
    \label{eq:Zk2} z_{k,2}(r, \Gammab) & \triangleq \mathbbm{1} \bigg[ \log_{2} \bigg( 1 + \bigg| \sum_{j \in \setK \setminus \{ k \}} z_{j,1}(r,\Gammab) \sqrt{\xi_j} h_{jk} \bigg|^2 \bigg) \geq r \bigg]
\end{align}
which are equal to $1$ if the \gls{ue} successfully decodes in the first and in the second phase, respectively, and to $0$ otherwise. Hence, the probabilities that \gls{ue}~$k$ successfully decodes in the first and in the second phase are given by
\begin{align}
    \rmP_{k,1}(r,\Gammab) = z_{k,1}(r,\Gammab), \\
    \rmP_{k,2}(r,\Gammab) = z_{k,2}(r,\Gammab)
\end{align}
respectively: these stem from the fact that, with perfect \gls{csit}, the decodability of each \gls{ue} in each phase is deterministic. In this context, the average success probability in \eqref{eq:av_P} can be written as
\begin{align}
    \rmP_{\mathrm{A}}(r, \Gammab) = \frac{1}{K}\sum_{k\in \setK} \big(z_{k,1}(r,\Gammab) + \big(1-z_{k,1}(r,\Gammab)\big)z_{k,2}(r,\Gammab)\big).
\end{align}
On the other hand, the joint success probability in \eqref{eq:Prob_ph2J} becomes a product of binary variables, which is equal to $0$ if even a single \gls{ue} does not decode the message over the two phases: hence, it is not suited to accommodate any target outage in the case of perfect \gls{csit}. For this reason, in the rest of the section, we focus on maximizing the average multicast rate in \eqref{eq:R_av}.

%=========================================================================
\subsection{Multi-Antenna Multicasting (MAM) Algorithm} \label{subsec:D2D-MAM_bl}
%=========================================================================

Considering the single-phase baseline scheme described in Section~\ref{subsec:baseline}, problem~\eqref{eq:Problem} with $\mathrm{T} = \mathrm{A}$ and perfect \gls{csit} can be solved by selecting the best subset of $\setK$ with size $(1-\epsilon)K$ to be served by the \gls{bs} and computing the transmit covariance that maximizes the multicast rate over such a subset of \glspl{ue}.\footnote{Without loss of generality, one can assume that $\epsilon$ is chosen such that $(1-\epsilon) K$ is an integer number.} Note that, in this case, the outage constraint in \eqref{eq:Problem} can be simply expressed as $\sum_{k \in \setK} z_{k,1}(r, \Gammab) \geq (1-\epsilon)K$. While this problem formulation is also novel, it mainly serves as a benchmark to demonstrate the gains obtained by the adding a second phase of \gls{ue} cooperation enabled by \gls{d2d} links in Section~\ref{sec:NR}. However, the problem of deriving the optimal \gls{ue} selection strategy is NP-hard since it requires to evaluate all possible subsets of $\setK$ with size $(1-\epsilon)K$. To reduce the complexity, we build on the intuition described in the following lemma to derive a suboptimal \gls{ue} selection scheme. \vspace{-1mm}

\begin{lemma} \label{lem:baseline}
For a class of channels satisfying $\Exp[\h_k \h_k^{\herm} ] = \gamma_k \I_M$, $\forall k \in \setK$, which includes \eqref{eq:channel}, the optimal \gls{ue} selection strategy with statistical channel knowledge is the one choosing the $(1-\epsilon)K$ \glspl{ue} with the highest average channel power gains among $\{\gamma_{k}\}_{k \in \setK}$.
\end{lemma} \vspace{-1mm}

\begin{IEEEproof}
If $\{\gamma_k > 0\}_{k \in \setK}$ are known at the \gls{bs}, we have
\begin{align}
    \nonumber & \max_{\setU \subset \setK \; : \; |\setU | = (1 - \epsilon) K} \Exp \Big[ \max_{\Gammab \succeq \0 \; : \; \tr(\Gammab) \leq 1} \min_{k \in \setU} \h_k^{\herm} \Gammab \h_k \Big] \\
    \label{eq:lem1_proof1} & \leq \max_{\setU \subset \setK \; : \; |\setU| = (1 - \epsilon) K} \; \max_{\Gammab \succeq \0 \; : \; \tr(\Gammab) \leq 1} \min_{k \in \setU} \Exp[\h_k^{\herm} \Gammab \h_k ] \\
    & = \max_{\setU \subset \setK \; : \; |\setU|=(1-\epsilon)K} \; \max_{\Gammab \succeq \0 \; : \; \tr(\Gammab) \leq 1} \min_{k \in \setU} \tr \big( \Gammab \Exp[\h_k \h_k^{\herm}] \big) \\
    \label{eq:lem1_proof2} & = \max_{\setU \subset \setK \; : \; |\setU| = (1 - \epsilon) K} \min_{k \in \setU} \gamma_k
\end{align}
where \eqref{eq:lem1_proof1} follows from the concavity of $\min_{k \in \setU} \h_k^{\herm} \Gammab \h_k$ and \eqref{eq:lem1_proof2} is due to the fact that the optimal $\Gammab$ satisfies $\tr(\Gammab) = 1$. Finally, the solution presented in the lemma readily follows from \eqref{eq:lem1_proof2}.
\end{IEEEproof} \vspace{1mm}

Lemma~\ref{lem:baseline} states that, if the channels can be ordered statistically based on the average channel power gains $\{\gamma_{k}\}_{k \in \setK}$, the exhaustive search over all possible subsets of $\setK$ with size $(1-\epsilon)K$ reduces to choosing the $(1-\epsilon) K$ \glspl{ue} with the highest $\gamma_{k}$. Motivated by this observation, we thus propose to apply such a \gls{ue} selection strategy to the case of perfect \gls{csit} and obtain the \textit{multi-antenna multicasting (MAM) algorithm}. More specifically, we build $\setU \subset \setK$ by selecting the $(1-\epsilon) K$ \glspl{ue} with the highest channel power gain $\| \h_k \|^2$ and compute the transmit covariance that achieves the multicast capacity over $\setU$, i.e.,
\begin{align}\label{eq:Sigma_bsl}
    \Gammab_{1} = \argmax_{\Gammab \succeq \mathbf{0} \; : \; \mathrm{tr}(\Gammab) \leq 1} \min_{k\in \setU} \h_{k}^{\herm}\Gammab\h_{k}.
\end{align}
Since the whole time resource is dedicated to the first phase, the resulting average multicast rate is given by
\begin{align}\label{eq:r_bsl}
    r_{1} = \log_2 \Big( 1 + \xi_0  \min_{k\in \setU} \h_{k}^{\herm}\Gammab_{1}\h_{k} \Big).
\end{align}

%=========================================================================
\subsection{\gls{d2d}-Aided Multi-Antenna Multicasting (\nameP{}) Algorithm} \label{subsec:D2D-MAM}
%=========================================================================

To solve problem~\eqref{eq:Problem} with $\mathrm{T} = \mathrm{A}$ and perfect \gls{csit}, we resort to the alternating optimization of the multicast rate $r$ and the transmit covariance $\Gammab$. In this respect, we propose an efficient iterative algorithm whose goal is to serve a subset of \glspl{ue} (which are suitably selected by means of precoding at the \gls{bs}) in the first phase such that the multicast rate is maximized. At each iteration $n$, the transmit covariance $\Gammab^{(n)}$ that achieves the multicast capacity over a predetermined subset $\setU^{(n-1)} \subset \setK$ is computed (see \eqref{eq:P_multicast}--\eqref{eq:P_multicast1}). Then, the multicast rate $r^{(n)}$ is obtained as the maximum rate that guarantees the outage constraint over the two phases given the transmit covariance computed in the previous step, i.e., such that $\rmP_{\mathrm{A}}(r^{(n)}, \Gammab^{(n)}) \geq 1 - \epsilon$. The new $r^{(n)}$ yields an updated $\setU^{(n)}$ of \glspl{ue} that are able to decode in the first phase and, therefore, an improved transmit covariance can be obtained by optimizing over $\setU^{(n)}$. This procedure is iterated until the multicast rate converges. The proposed algorithm is referred to as \textit{\gls{d2d}-aided multi-antenna multicasting (\nameP{}) algorithm} and is formally described in Algorithm~\ref{alg:A1}. The \nameP{} algorithm has the key advantage of not requiring any tuning parameter selection. Furthermore, it converges to a local optimum of problem~\eqref{eq:Problem} with $\mathrm{T} = \mathrm{A}$, as formalized in the following theorem. \vspace{-1mm}

\begin{figure}[t!]
\begin{algorithm}[H]
\begin{algorithmic}
\begin{spacing}{1.25}
\STATE \hspace{-4mm} \textbf{Data:} Direct channels $\{\h_k\}_{k \in \setK}$ and \gls{d2d} channels $\{h_{jk}\}_{k,j \in \setK}$. Fix $\setU^{(0)} = \setK$ and $n=1$.
\begin{itemize}[leftmargin=12mm]
    \item[\texttt{(S.1)}] Optimize the transmit covariance as
    \begin{align*}
        \Gammab^{(n)} = \argmax_{\Gammab \succeq \mathbf{0} \; : \; \mathrm{tr}(\Gammab) \leq 1} \min_{k \in \setU^{(n-1)}} \h_{k}^{\herm}\Gammab\h_{k}.
    \end{align*}
    \item[\texttt{(S.2)}] Maximize the multicast rate as
    \begin{align*}
        r^{(n)} = \max \big\{r : \rmP_{\mathrm{A}}(r,\Gammab^{(n)}) = 1 - \epsilon \big\}.
    \end{align*}
    \item[\texttt{(S.3)}] Update the subset of \glspl{ue} successfully decoding in the first phase as
    \begin{align*}
        \setU^{(n)} = \big\{ k : \log_{2} (1 + \xi_0 \h_{k}^{\herm} \Gammab^{(n)} \h_{k}) \geq r^{(n)} \big\}.
    \end{align*}
    \item[\texttt{(S.4)}] \texttt{If} $r^{(n)} = r^{(n-1)}$: fix $\Gammab = \Gammab^{(n)}$ and $r = r^{(n)}$; \texttt{Stop}.
    \item[] \texttt{Else}: $n \leftarrow n+1$; \texttt{Go to (S.1)}.
\end{itemize}
\vspace{-5mm}
\end{spacing}
\end{algorithmic}
\caption{(\nameP{})} \label{alg:A1}
\end{algorithm}
\vspace{-5mm}
\end{figure}

\begin{theorem}\label{thm:alg1}
The \nameP{} algorithm converges to a local optimum of problem~\eqref{eq:Problem} with $\mathrm{T} = \mathrm{A}$.
\end{theorem} \vspace{-1mm}

\begin{IEEEproof}
Since step~(S.1) of Algorithm~\ref{alg:A1} optimizes $\Gammab^{(n)}$ over $\setU^{(n-1)}$, we have
\begin{align} \label{eq:Sigma_n}
    \min_{k\in \setU^{(n-1)}} \h_k^{\herm}\Gammab^{(n)}\h_k \geq \min_{k\in \setU^{(n-1)}} \h_k^{\herm}\Gammab^{(n-1)}\h_k
\end{align}
i.e., the minimum rate achievable by the \glspl{ue} in $\setU^{(n-1)}$ increases with the new transmit covariance $\Gammab^{(n)}$. Furthermore, at each iteration $n$ of the \nameP{} algorithm, the following holds:
\begin{align}
    r^{(n)} & \geq \log_2 \Big(1 + \rho \min_{k\in \setU^{(n-1)}} \h_k^{\herm}\Gammab^{(n)}\h_k \Big) \label{eq:r_n1} \\
    & \geq \log_2 \Big( 1 + \rho \min_{k\in \setU^{(n-1)}} \h_k^{\herm}\Gammab^{(n-1)}\h_k \Big) \label{eq:r_n2} \\
    & \geq r^{(n-1)} \label{eq:r_n3}
\end{align}
where \eqref{eq:r_n1} follows from step~(S.2) of Algorithm~\ref{alg:A1} (by which it is possible to increase the multicast rate as long as the outage constraint is guaranteed), \eqref{eq:r_n2} is a direct consequence of \eqref{eq:Sigma_n}, and \eqref{eq:r_n3} stems from the fact that $\setU^{(n-1)}$ contains the \glspl{ue} whose achievable rate in the first phase is at least $r^{(n-1)}$. Hence, the multicast rate cannot decrease between consecutive iterations. Finally, if $\setU^{(n)}=\setU^{(n-1)}$, then it is not possible to further increase the multicast rate, i.e., $r^{(n)}=r^{(n-1)}$, which implies that convergence is reached.
\end{IEEEproof} \vspace{1mm}

Regarding the optimization of the multicast rate in step~(S.2) of Algorithm~\ref{alg:A1}, we have
\begin{align}
    r^{(n)} \in \Big[ r^{(n-1)}, \log_2 \Big( 1+\rho \max_{k \in \setU^{(n-1)}} \h_k^{\herm}\Gammab^{(n)}\h_k \Big) \Big]
\end{align}
where the lower bound follows from Theorem~\ref{thm:alg1} and the upper bound is necessary to guarantee that at least one \gls{ue} is served in the first phase: thus, $r^{(n)}$ can be efficiently computed by means of bisection over the above interval. Accordingly, every iteration of the \nameP{} algorithm requires the solution of a convex problem in step~(S.1) and a linear search in step~(S.2); in addition, for the settings considered for our simulations in Section~\ref{sec:NR}, convergence is reached after a small number of iterations. Hence, the \nameP{} algorithm provides a locally optimal solution of problem~\eqref{eq:Problem} with $\mathrm{T} = \mathrm{A}$ with very low complexity.

%=========================================================================
\section{\gls{d2d}-Aided Multi-Antenna Multicasting with Statistical \gls{csit}} \label{sec:statCSI}
%=========================================================================

In this section, we consider the case where only the \gls{ue} locations are known at the \gls{bs}. From this information, the \gls{bs} can extract long-term statistics such as the average channel power gains of both the direct channels, i.e., $\{\gamma_k\}_{k \in \setK}$, and the \gls{d2d} channels, i.e., $\{\gamma_{jk}\}_{k,j \in \setK}$, together with the steering angles $\{\theta_k\}_{k \in \setK}$. On the other hand, the \gls{bs} has no knowledge of the small-scale fading coefficients, i.e., $\{\eta_k\}_{k \in \setK}$ and $\{\eta_{jk}\}_{k,j \in \setK}$. Under statistical \gls{csit}, we characterize the service reliability in terms of the joint success probability in \eqref{eq:Prob_ph2J} and, accordingly, we maximize the outage multicast rate in \eqref{eq:R_out}. To alleviate the task of dealing with the involved expression of the joint success probability, we derive its deterministic equivalent in the following proposition. \vspace{-1mm}

\begin{proposition} \label{eq:pjoint2}
Assuming that all (direct and \gls{d2d}) channels are independent, we have
\begin{align}
    \rmP_{\mathrm{J}}(r,\Gammab) \underset{K\to \infty}{\overset{\Pr}{\to}} \bar{\rmP}_{\mathrm{J}}(r,\Gammab)
\end{align} 
where
\begin{align} \label{eq:pjoint2_K}
    \bar{\rmP}_{\mathrm{J}}(r,\Gammab) & \triangleq \mathrm{exp}\bigg(-\sum_{k\in \setK}\frac{(2^r-1) \big( 1-\rmP_{k,1}(r,\Gammab) \big)}{\sum_{j\in \setK \setminus\{k\}}\rmP_{j,1}(r,\Gammab)\gamma_{jk}\xi_j}\bigg)
\end{align}
is the deterministic equivalent of $\rmP_{\mathrm{J}}(r,\Gammab)$ in \eqref{eq:Prob_ph2J}.
\end{proposition} \vspace{-1mm}

\begin{IEEEproof}
The proof follows similar steps as the proof of \cite[Thm.~4]{Sant20} and is thus omitted.
\end{IEEEproof} \vspace{1mm}

\noindent Note that, with statistical \gls{csit}, the probability that \gls{ue}~$k$ successfully decodes in the first phase is given by
\begin{align}
    \rmP_{k,1}(r, \Gammab) & = \Pr \big[ z_{k,1}(r,\Gammab) = 1 \big] \\
    & = \Pr \big[ \log_2(1 + \xi_0 \gamma_{k} |\eta_k|^2 \a^{\herm}_k \Gammab \a_k) \geq r \big] \\
    & = \exp \bigg( - \frac{2^{r}-1}{\xi_0 \gamma_{k} \a^{\herm}_k \Gammab \a_k} \bigg) \label{eq:pk1}.
\end{align}
with $z_{k,1}(r,\Gammab)$ defined in \eqref{eq:Zk1} and where \eqref{eq:pk1} follows from the exponential distribution of $|\eta_{k}|^{2}$. In the rest of the section, we replace $\rmP_{\mathrm{J}}(r,\Gammab)$ with its deterministic equivalent $\bar{\rmP}_{\mathrm{J}}(r,\Gammab)$ in \eqref{eq:pjoint2_K}.

%=========================================================================
\subsection{Statistical Multi-Antenna Multicasting (SMAM) Algorithm} \label{sec:D2D-SMAM_bl}
%=========================================================================

Considering the single-phase baseline scheme described in Section~\ref{subsec:baseline}, problem~\eqref{eq:Problem} with $\mathrm{T} = \mathrm{J}$ and statistical \gls{csit} can be solved by computing the transmit covariance that maximizes the outage multicast rate. Note that, in this case, the outage constraint in \eqref{eq:Problem} can be simply expressed as $\prod_{k \in \setK} \rmP_{k,1}(r,\Gammab) \geq 1-\epsilon$. Since this problem is convex in $\Gammab$ for a fixed $r$ and vice versa, we decouple the optimization over the two variables in the following way. For a given transmit covariance $\Gammab_1$, the outage multicast rate, denoted in this context by $R_{\mathrm{O},1}(r_{1},\Gammab_1)$, is maximized when the outage constraint is satisfied with equality, leading to
\begin{align} \label{eq:r_opt_bl}
    R_{\mathrm{O},1}(r_{1},\Gammab_1) = \log_2\bigg(1+\xi_0 \log\bigg(\frac{1}{1-\epsilon}\bigg)\bigg(\sum_{k\in \setK}\frac{1}{\gamma_k \a_k^{\herm}\Gammab_1\a_k}\bigg)^{-1}\bigg).
\end{align}
Then, the optimal transmit covariance is obtained by solving
\begin{align} \label{eq:p1_bl}
    \begin{array}{cl}
    \displaystyle \min_{\Gammab_1\succeq \0} & \displaystyle \sum_{k\in \setK}\frac{1}{\gamma_k\a_k^{\herm}\Gammab_1\a_k} \\
    \mathrm{s.t.}& \tr(\Gammab_1)\leq 1
    \end{array}
\end{align}
by means of semidefinite programming. As in Section~\ref{subsec:D2D-MAM_bl}, this problem formulation mainly serves for the comparative purposes in Section~\ref{sec:NR}. The resulting algorithm is referred to as \textit{statistical multi-antenna multicasting (SMAM) algorithm}.

The following proposition derives a tractable expression of $\Gammab_1$ and will be useful in the next section. \vspace{-1mm}

\begin{proposition}\label{prop:aorth}
Assume that $\setK$ consists of $M$ \glspl{ue} exhibiting mutually orthogonal array responses, i.e.,
\begin{align} \label{eq:aorth}
    \sum_{k\in \setK} \a_k\a_k^{\herm} = M \I_M.
\end{align}
Then, the optimal transmit covariance for problem~\eqref{eq:p1_bl} can be written in closed form as
\begin{align} \label{eq:gamma_aorth}
    \Gammab_{1} = \frac{1}{M \nu_{\setK}} \sum_{k\in \setK} \frac{1}{\sqrt{\gamma_k}} \a_k\a_k^{\herm}
\end{align}
with $\nu_{\setK} \triangleq \sum_{k\in \setK} \frac{1}{\sqrt{\gamma_k}}$.
\end{proposition} \vspace{-1mm}

\begin{IEEEproof}
See Appendix~\ref{ap:Gamma_orth}.
\end{IEEEproof} \vspace{1mm}

\noindent A set of array response vectors satisfying \eqref{eq:aorth} can be obtained as the columns of the $M$-dimensional discrete Fourier transform (DFT) matrix or, alternatively, it can be constructed along specific virtual angles as described in \cite{Saye02}.

%=========================================================================
\subsection{\gls{d2d}-Aided Statistical Multi-Antenna Multicasting (\nameS{}) Algorithm}
%=========================================================================

To solve problem \eqref{eq:Problem} with $\mathrm{T} = \mathrm{J}$ and statistical \gls{csit}, we use the deterministic equivalent derived in Proposition~\ref{eq:pjoint2} and, to further reduce the complexity, we decouple the optimization across the two phases in the following way. First, we carefully select a subset $\setU \subset \setK$ of \glspl{ue} with favorable statistical properties to be served in the first phase by the \gls{bs}. In particular, assuming large $K$ and uniform \gls{ue} distribution in the angular domain, we build on Proposition~\ref{prop:aorth} and construct $\setU$ by selecting $M$ \glspl{ue} satisfying the condition in \eqref{eq:aorth}:\footnote{Since $K$ is large, we assume that it is always possible to select $M$ \glspl{ue} whose steering angles satisfy \eqref{eq:a_orth}.} by doing so, the \gls{bs} spreads its transmit power along a set of orthogonal directions spanning the whole angular domain. In this setting, the transmit covariance that maximizes the multicast rate over $\setU$ is given by $\Gammab_1$ in \eqref{eq:gamma_aorth}. Next, we fix the joint success probability in the first phase over $\setU$ to a given value $1-\epsilon_1$ and obtain the corresponding multicast rate $r(\epsilon_1)$ from \eqref{eq:r_opt_bl}. Finally, we optimize $\epsilon_1$ in order to obtain the desired joint success probability $1-\epsilon$ over the two phases.

Let us first focus on maximizing the outage multicast rate over $\setU$ in the first phase, i.e.,
\begin{align}\label{eq:P00}
    \begin{array}{cl}
    \displaystyle \max_{r(\epsilon_1)>0, \Gammab \succeq \0} & \displaystyle  r(\epsilon_1) \\
    \mathrm{s.t.}& \mathrm{tr}(\Gammab) \leq 1, \\
    & \displaystyle \exp \bigg( - \sum_{k \in \setU} \frac{2^{r(\epsilon_1)}-1}{\xi_0 \gamma_{k} \a^{\herm}_k \Gammab \a_k} \bigg) \geq 1 - \epsilon_1.
    \end{array}
\end{align}
Since the outage constraint is convex in $\Gammab$, we can solve problem~\eqref{eq:P00} by decoupling the optimization of $r(\epsilon_1)$ and $\Gammab$. Letting the outage constraint be satisfied with equality, we have that the multicast rate becomes
\begin{align}\label{eq:r0}
    r(\epsilon_1) = \log_{2} \bigg( 1 + \xi_0 \log \bigg( \frac{1}{1-\epsilon_1} \bigg) \bigg( \sum_{k \in \setU} \frac{1}{\gamma_{k} \a^{\herm}_k \Gammab \a_k} \bigg)^{-1} \bigg)
\end{align}
and problem~\eqref{eq:P00} reduces to finding the transmit covariance $\Gammab$ by solving
\begin{align} \label{eq:P0}
    \begin{array}{cl}
    \displaystyle \min_{\Gammab \succeq \0} & \displaystyle \sum_{k\in \setU} \frac{1}{\gamma_k \a_k^{\herm} \Gammab \a_k} \\
    \mathrm{s.t.} & \tr(\Gammab) \leq 1.
    \end{array}
\end{align}
From Proposition~\ref{prop:aorth}, the transmit covariance resulting from the above problem is known to have a simple closed-form expression when $|\setU| = M$ and the \glspl{ue} in $\setU$ exhibit orthogonal array response vectors, i.e.,
\begin{align}\label{eq:a_orth}
    \sum_{k \in \setU} \a_k\a_k^{\herm} = M \I_M.
\end{align}
Since $K$ is large, we assume that it is always possible to build $\setU$ by selecting $M$ \glspl{ue} satisfying the condition in \eqref{eq:a_orth}. In this case, the optimal transmit covariance is given by 
\begin{align} \label{eq:448}
    \Gammab = \sum_{j \in \setU} w_j \a_j \a_j^{\herm} 
\end{align}
with weights given by
\begin{align} \label{eq:w_opt}
    w_j \triangleq \frac{1}{M\bar{\gamma}_{\setU} } \frac{1}{\sqrt{\gamma_j}} \quad \forall j \in \setU
\end{align}
and where we have defined $\bar{\gamma}_{\setU} \triangleq \sum_{k \in \setU} \frac{1}{\sqrt{\gamma_{k}}}$. Finally, plugging \eqref{eq:448} and \eqref{eq:w_opt} into \eqref{eq:r0}, we obtain
\begin{align} \label{eq:r_orth}
    r(\epsilon_1) & = \log_2 \bigg( 1 + \xi_0 \log \bigg( \frac{1}{1-\epsilon_1} \bigg) \frac{M}{\bar{\gamma_{\setU}}^2} \bigg).
\end{align}

Let us now focus on deriving $\epsilon_1$ that achieves the desired joint success probability $1-\epsilon$ over the two phases. This can be done by solving the following expression for $\epsilon_1 \in [0,1)$ (e.g., by means of bisection):
\begin{align}\label{eq:e1_deter}
    (2^{r(\epsilon_1)}-1)\sum_{k\in \setK}\frac{1-\mathrm{exp}\big(-\frac{2^{r(\epsilon_1)}-1}{\xi_0\gamma_k\a_k^{\herm}\Gammab\a_k}\big)}{\sum_{j \in \setK \setminus \{k\}}\mathrm{exp}\big(-\frac{2^{r(\epsilon_1)}-1}{\xi_0\gamma_k\a_k^{\herm}\Gammab\a_k}\big)\gamma_{jk}\xi_j} \leq \log\bigg(\frac{1}{1-\epsilon}\bigg).
\end{align}
The proposed algorithm is referred to as \textit{\gls{d2d}-aided statistical multi-antenna multicasting (\nameS{}) algorithm} and is formally described in Algorithm~\ref{alg:A2}. In the next section, we illustrate a possible way to derive an approximation of the optimal $\epsilon_1$.

\begin{figure}[t!]
\begin{algorithm}[H]
\begin{algorithmic}
\begin{spacing}{1.25}
\STATE \hspace{-4mm} \textbf{Data:} Build $\setU$ by selecting $M$ \glspl{ue} such that \eqref{eq:a_orth} holds.
\begin{itemize}[leftmargin=12mm]
    \item[\texttt{(S.1)}] Compute the transmit covariance as in \eqref{eq:448} with weights given in \eqref{eq:w_opt}.
    \item[\texttt{(S.2)}] Find $\epsilon_1$ by solving \eqref{eq:e1_deter}.
    \item[\texttt{(S.3)}] Compute the multicast rate as in \eqref{eq:r_orth}.
\end{itemize}
\vspace{-5mm}
\end{spacing}
\end{algorithmic}
\caption{(\nameS{})} \label{alg:A2}
\end{algorithm}
\vspace{-5mm}
\end{figure}

%=========================================================================
\subsection{Asymptotic Behavior of the \nameS{} Algorithm} \label{subsec:Asympt}
%=========================================================================

Let us assume that $\epsilon \to 0$ and, consequently, that $\epsilon_1 \to 0$. By applying the Taylor approximation $\mathrm{exp}\left(-\frac{2^{r(\epsilon_1)-1}}{\xi_0\gamma_k\a_k^{\herm}\Gammab\a_k}\right)\approx 1-\frac{2^{r(\epsilon_1)-1}}{\xi_0\gamma_k\a_k^{\herm}\Gammab\a_k}$ to \eqref{eq:e1_deter}, we have
\begin{align}
    \epsilon_1 \underset{\epsilon\to 0}{\sim} 1 - \mathrm{exp} \Bigg( -\frac{\bar{\gamma}_{\setU}^2}{M\xi_0}\sqrt{\frac{\log\big(\frac{1}{1-\epsilon}\big)}{\sum_{k\in \setK}\frac{1}{\xi_0 \gamma_k \a_k^{\herm}\Gammab\a_k}\big(\sum_{j\in\setK\setminus\{k\}}\gamma_{jk}\xi_j\big)^{-1}}} \Bigg)
\end{align}
and, hence
\begin{align}
    \label{eq:r_approx_taylor1} r & \underset{\epsilon\to 0}{\sim} \frac{1}{2}\log_2 \Bigg( 1 + \sqrt{\frac{\xi_0\log\big(\frac{1}{1-\epsilon}\big)}{\sum_{k\in \setK}\frac{1}{ \gamma_k \a_k^{\herm}\Gammab\a_k}\big(\sum_{j\in\setK\setminus\{k\}}\gamma_{jk}\xi_j\big)^{-1}}} \Bigg) \\
    \label{eq:r_approx_taylor2} & \ \triangleq \ \tilde{r}.
\end{align}
Now, assume that $d_k \in [R_{\min}, R_{\max}]$, $\forall k \in \setK$, where $R_{\min}$ and $R_{\max}$ denote the minimum and maximum distance, respectively, between each \gls{ue} and the \gls{bs}. It follows that the average channel power gains can be bounded as
\begin{align}
    \label{eq:gamma_kworst} \gamma_k & \in [R_{\max}^{-\beta}, R_{\min}^{-\alpha}], \hspace{2.42cm} \forall k \in \setK, \\
    \label{eq:gamma_jkworst} \gamma_{jk} & \in \big[ (2R_{\max})^{-\beta}, (2R_{\min})^{-\alpha}], \quad \forall k,j \in \setK.
\end{align}
In this setting, we have
\begin{align}
    \a_k^{\herm}\Gammab\a_k & = \frac{1}{M \bar{\gamma}_{\setU}} \sum_{j\in \setU} \frac{1}{\sqrt{\gamma_j}} |\a_k^{\herm} \a_{j}|^{2} \label{eq:sum_2terms} \\
    & \geq \frac{M}{\bar{\gamma}_{\setU}}R_{\min}^{\alpha/2} \label{eq:snr_eff_bound}
\end{align}
where \eqref{eq:snr_eff_bound} follows from assuming that all the \glspl{ue} in $\setU$ are at distance $R_{\min}$ from the \gls{bs}, i.e., $\{ \gamma_{j} = R_{\min}^{-\alpha} \}_{j \in \setU}$. Hence, we have that $\tilde{r}$ defined in \eqref{eq:r_approx_taylor1}--\eqref{eq:r_approx_taylor2} can be lower bounded as
\begin{align} \label{eq:r_lower}
    \tilde{r} \geq \log_2 \Bigg( 1+\sqrt{\frac{\xi_0\xi_{\ue}(K-1)M\log\big(\frac{1}{1-\epsilon}\big) R_{\mathrm{min}}^{\alpha/2}}{\bar{\gamma}_{\setU}^2(2R_{\mathrm{max}})^{\beta}R_{\mathrm{min}}^{\alpha/2} + 2^{\beta}\bar{\gamma}_{\setU}R_{\mathrm{max}}^{2\beta}(K-M)}} \Bigg)
\end{align}
where, for simplicity, we have assumed that $\{ \xi_{k} = \xi_{\ue} \}_{k \in \setK}$ (i.e., all the \glspl{ue} have the same transmit SNR in the second phase). Finally, we consider the asymptotic behavior of $\tilde{r}$ in the case where both $K$ and $M$ increase with fixed ratio $c \triangleq \frac{K}{M} >1$ as well as in the case where $K$ increases for a fixed $M$. Hence, \eqref{eq:r_lower} behaves as
\begin{align}
    \tilde{r} \underset{K \to \infty}{\to} \begin{cases}
    \log_2 \bigg( 1+\sqrt{\frac{\xi_0\xi_{\ue}M\log\big(\frac{1}{1-\epsilon}\big) R_{\mathrm{min}}^{\alpha/2}}{2^{\beta}\bar{\gamma}_{\setU}R_{\mathrm{max}}^{2\beta}}} \bigg) & \textrm{for fixed}~M \\
    \log_2 \bigg( 1+\sqrt{\frac{\xi_0\xi_{\ue}\log\big(\frac{1}{1-\epsilon}\big) R_{\mathrm{min}}^{\alpha/2}}{2^{\beta}\bar{\gamma}_{\setU}R_{\mathrm{max}}^{2\beta}(c-1)}K} \bigg) & \textrm{for}~M=\frac{K}{c},~\textrm{with}~c>1
    \end{cases}
\end{align}
which is non-vanishing in the first case as in \cite{Sant20} and increasing as $\log_{2} \big( 1 + \sqrt{K} \big)$ in the second case.

%=========================================================================
\section{\gls{d2d}-Aided Multi-Antenna Multicasting with Topological \gls{csit}} \label{sec:continuous}
%=========================================================================

In this section, we consider the case where only the map of the network area, i.e., the location and size of the obstacles (such as buildings) within its coverage area, and the \gls{pdf} of the \gls{ue} locations are known at the \gls{bs}. Such \gls{pdf} can be obtained on the basis of the city map and long-term information on the traffic distribution. This setting describes a scenario with a high density of \glspl{ue} (e.g, cars or terminals) where it may not be feasible to design a precoding solution that adapts instantaneously to the channels or, in the longer term, to the channel statistics. In this case, it is meaningful to derive the precoding strategy at the \gls{bs} based solely on the network topology and on the \gls{ue} distribution.

First, we slightly adapt the channel model described in Section~\ref{subsec:CM} to express all the parameters as functions of the possible \gls{ue} locations within the map. Let $\setA \subset \Real^2$ denote the continuous set of points representing the network area and let $\p = (\theta,\rho)$ be a random variable denoting a possible position within $\setA$ in which a \gls{ue} can be located, where $\theta$ and $\rho$ represent the steering angle and the distance from the \gls{bs}, respectively. In this setting, we use $f(\p)$ to denote the \gls{pdf} of the \gls{ue} locations, which describes the probability of finding a \gls{ue} in the position identified by $\p$. Focusing on the first phase, let us write the direct channel to position $\p$ as (cf. \eqref{eq:channel})
\begin{align} \label{eq:h(p)}
    \h(\p) = \eta\sqrt{\gamma(\p)}\, \a(\theta)
\end{align}
where $\eta\sim \mathcal{CN}(0,1)$ is the small-scale fading coefficient, $\gamma(\p)$ is the average channel power gain at position $\p$, and $\a(\theta)$ is the array response vector at the \gls{bs} for the steering angle $\theta$. Here, we have $\gamma(\p) = \rho^{-\alpha}$ in case of \gls{los} conditions and $\gamma(\p) = \rho^{-\beta}$ in case of \gls{nlos} conditions. The receive SNR at position $\p$ in the first phase can be expressed as
\begin{align} \label{eq:SNR_topol}
    \mathrm{SNR}_1(\p,\Gammab) \triangleq |\eta|^2 \gamma(\p) \xi_0 \a^{\herm}(\theta)\Gammab\a(\theta).
\end{align}
Note that, if position $\p$ falls within the area occupied by an obstacle (e.g., a building), the corresponding receive SNR is zero. Hence, the probability that a \gls{ue} located at position $\p$ successfully decodes in the first phase is given by
\begin{align}\label{eq:P_1_topol}
    \rmP_1(\p,r,\Gammab) & \triangleq \Pr \big[ \log_2 \big( 1+\mathrm{SNR}_1(\p,\Gammab) \big) \geq r \big] \\
    & = \mathrm{exp}\bigg(-\frac{(2^r-1)}{\gamma(\p)\xi_0 \a^{\herm}(\theta)\Gammab\a(\theta)}\bigg).
\end{align}
Focusing on the second phase, let us write the \gls{d2d} channel between positions $\p$ and $\p^{\prime}$ as (cf.~\eqref{eq:channel_d2d})
\begin{align} \label{eq:h(p,p')}
    h(\p,\p^{\prime}) = \eta\sqrt{\gamma(\p,\p^{\prime})}
\end{align}
where $\gamma(\p,\p^{\prime})$ is the average channel power gain. Here, we have $\gamma(\p,\p^{\prime}) = d(\p,\p^{\prime})^{-\alpha}$ in case of \gls{los} conditions and $\gamma(\p,\p^{\prime}) = d(\p,\p^{\prime})^{-\beta}$ in case of \gls{nlos} conditions, where $d(\p,\p^{\prime})$ denotes the distance between positions $\p$ and $\p^{\prime}$. For simplicity, let us assume that all the \glspl{ue} in any position within $\setA$ have the same transmit SNR $\xi_{\ue}$ in the second phase. Furthermore, let $\setU \subset \setA$ be the subset of positions where a potential \gls{ue} could successfully decode in the first phase. Hence, the probability that a \gls{ue} located at position $\p$ successfully decodes the message in the second phase is given by
\begin{align} \label{eq:P_2_topol}
    \rmP_2(\p,r,\Gammab) & \triangleq \Pr\Bigg[\log_2\Bigg(1+\bigg\rvert\underset{\setU}{\int}\sqrt{\xi_{\ue}} f(\p^{\prime}) h(\p,\p^{\prime}) \diff \p^{\prime} \bigg\rvert^2\bigg)\geq r\bigg] \\
    & = \Exp\bigg[\mathrm{exp}\bigg(-\frac{2^r-1}{\xi_{\ue}\underset{\setU}{\int}f(\p^{\prime})\gamma(\p,\p{\prime}) \diff \p^{\prime}}\bigg)\bigg]
\end{align}
where the expectation is over all the possible combinations of $\setU$.

In the context of topological \gls{csit}, the average success probability in \eqref{eq:av_P} can be written as
\begin{align} \label{eq:P_av_succ_cont2}
    \rmP_{\mathrm{A}}(r, \Gammab) = \underset{\mathcal{A}}{\int} f(\p)\big(\rmP_1(\p,r,\Gammab) + (1-\rmP_1(\p,r,\Gammab))\rmP_2(\p,r,\Gammab)\big) \diff \p.
\end{align}
On the other hand, the joint success probability in \eqref{eq:Prob_ph2J} turns out to be impractical when $\setA$ is connected, i.e., when the network area contains infinite points. For this reason, in the rest of the section, we focus on maximizing the average multicast rate in \eqref{eq:R_av}. Since \eqref{eq:P_av_succ_cont2} is quite difficult to handle even for simple \gls{ue} distribution models (e.g., uniform), in the next section, we detail a heuristic approach to maximize the average multicast rate based on the Monte Carlo sampling of $f(\p)$.

\begin{figure}[t!]
\begin{algorithm}[H]
\begin{algorithmic}
\begin{spacing}{1.25}
\STATE \hspace{-4mm} \textbf{Data:} Map of the network area and \gls{pdf} of the \gls{ue} locations $f(\p)$. Fix $l=1$.
\STATE \hspace{-4mm} \texttt{For} $l=1,\ldots,L$:
\begin{itemize}[leftmargin=12mm]
    \item[\texttt{(S.1)}] Generate $T$ test points together with the corresponding direct and \gls{d2d} channels according to \eqref{eq:h(p)} and \eqref{eq:h(p,p')}, respectively.
    \item[\texttt{(S.2)}] Execute Algorithm~\ref{alg:A1} with the channels generated in step \texttt{(S.1)} as input data and obtain the multicast rate $r^{(\ell)}$ and the transmit covariance $\Gammab^{(\ell)}$ as output data.
\end{itemize}
\STATE \hspace{-4mm} \texttt{End}
\begin{itemize}[leftmargin=12mm]
    \item[\texttt{(S.3)}] Fix $r = \frac{1}{L} \sum_{\ell = 1}^{L} r^{(\ell)}$ and $\Gammab = \frac{1}{L} \sum_{\ell = 1}^{L} \Gammab^{(\ell)}$.
\end{itemize}
\vspace{-5mm}
\end{spacing}
\end{algorithmic}
\caption{(\nameT{})} \label{alg:A3}
\end{algorithm}
\vspace{-5mm}
\end{figure}

%=========================================================================
\subsection{\gls{d2d}-Aided Topological Multi-Antenna Multicasting (\nameT{}) Algorithm}
%=========================================================================

To solve problem \eqref{eq:Problem} with $\mathrm{T} = \mathrm{A}$ and topological \gls{csit}, we resort to the Monte Carlo sampling of the \gls{pdf} of the \gls{ue} locations to generate a set of test points within the map and the corresponding artificial channels, which are subsequently used to run the \nameP{} algorithm described in Algorithm~\ref{alg:A1} (see Section~\ref{subsec:D2D-MAM}). More specifically, we produce $L$ batches of $T$ test points each, where $T$ is a random variable that describes the number of \glspl{ue} and whose distribution depends on $f(\p)$. For each batch $\ell$, we artificially generate the direct channels for each test point according to \eqref{eq:h(p)} as well as the \gls{d2d} channels for each pair of test points according to \eqref{eq:h(p,p')}. Then, such channels are used as input data to Algorithm~\ref{alg:A1}, which produces the multicast rate $r^{(\ell)}$ and the transmit covariance $\Gammab^{(\ell)}$ as output data. Finally, the final multicast rate and transmit covariance are obtained by averaging the output data of the $L$ batches, i.e., $r = \frac{1}{L} \sum_{\ell = 1}^{L} r^{(\ell)}$ and $\Gammab = \frac{1}{L} \sum_{\ell = 1}^{L} \Gammab^{(\ell)}$, which provides an approximate solutions to problem~\eqref{eq:Problem} with $\mathrm{T} = \mathrm{A}$. The proposed algorithm is referred to as \textit{\gls{d2d}-aided topological multi-antenna multicasting (\nameT{}) algorithm} and is formally described in Algorithm~\ref{alg:A3}. Evidently, evaluating more batches of test points allows to achieve a more precise representation of the long-term network statistics, which produces a more accurate result in terms of average success probability. Since the \nameT{}~algorithm involves $L$ instances of Algorithm~\ref{alg:A1}, its computational complexity may be quite high. However, it is worth observing that this procedure is based on slowly varying network statistics and needs to be updated only when the \gls{ue} distribution changes significantly. Therefore, it can be conveniently executed offline using a large value of $L$.

%=========================================================================
\section{Numerical Results} \label{sec:NR}
%=========================================================================

\begin{figure}[t!]
\centering
\includegraphics[scale=0.45]{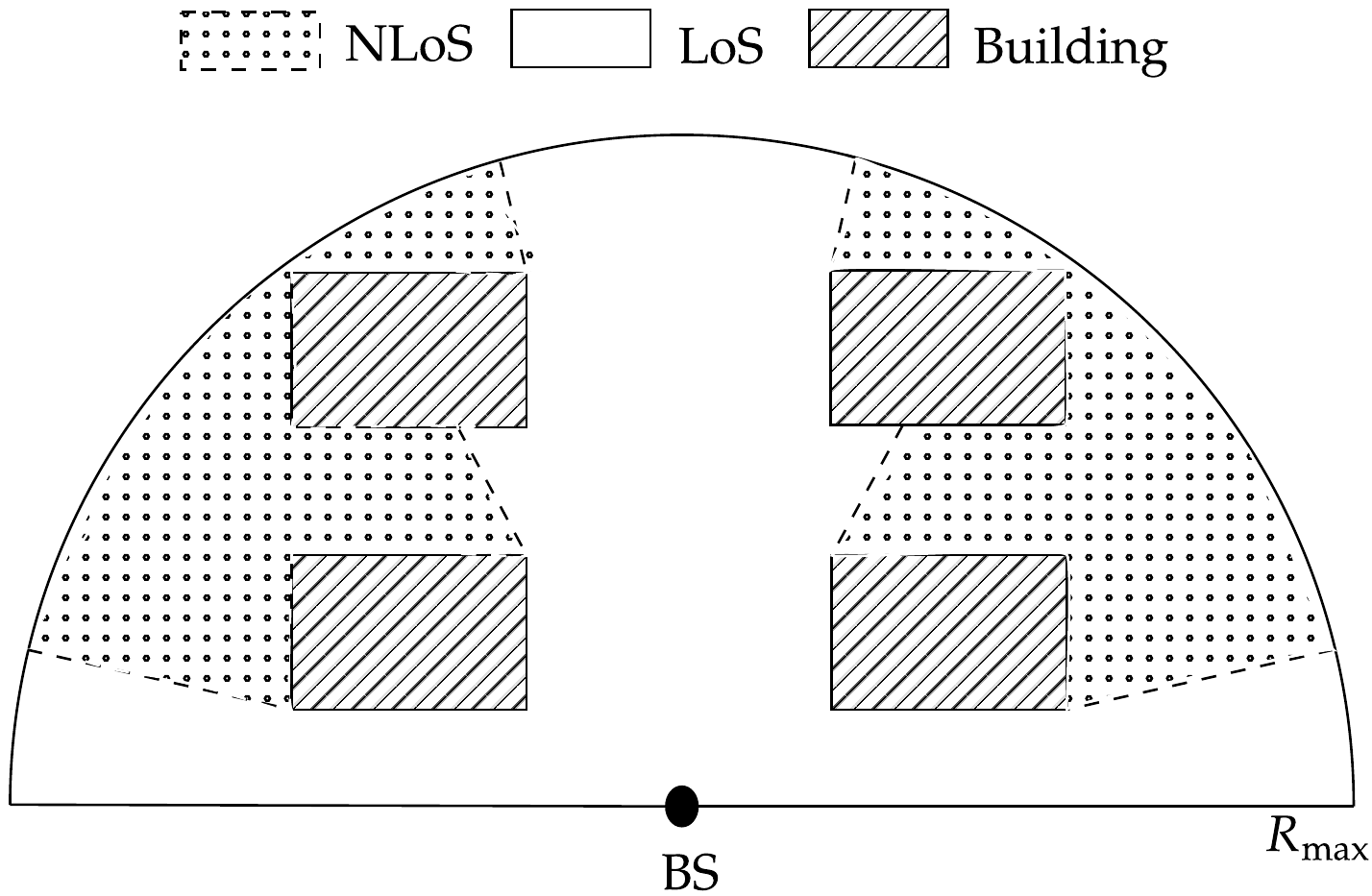}
\caption{Evaluation scenario: the white area and the dotted areas are in \gls{los} and \gls{nlos} conditions, respectively, whereas the \glspl{ue} are not admitted in the regions occupied by the buildings.} \label{fig:scenario}
\end{figure}

In this section, we present numerical results to validate the proposed algorithms in the three different \gls{csit} configurations, i.e., perfect \gls{csit} (described in Section~\ref{sec:perfCSI}), imperfect \gls{csit} (described in Section~\ref{sec:statCSI}), and topological \gls{csit} (described in Section~\ref{sec:continuous}). Unless otherwise stated, the considered network topology consists of a semicircular area with radius $R_{\mathrm{\max}} = 100$~m where four rectangular buildings are positioned in a Manhattan-like grid, as shown in Fig.~\ref{fig:scenario}. We assume that the \glspl{ue} are distributed uniformly within the network area with the exception of the regions occupied by the buildings and with a minimum distance from the \gls{bs} of $R_{\mathrm{min}} = 5$~m. The direct and \gls{d2d} links whose line of sight is obstructed by one or more buildings are considered to be in \gls{nlos} conditions both in the first and in the second phase. The \gls{los} and \gls{nlos} pathloss exponents are fixed to $\alpha = 2$ and $\beta = 4$, respectively. For simplicity, we assume that all the \glspl{ue} have the same transmit SNR in the second phase, i.e., $\{ \xi_{k} = \xi_{\ue} \}_{k \in \setK}$, and we set $\xi_0 = 30$~dB and $\xi_{\ue} = 20$~dB. Moreover, unless otherwise stated, the \gls{bs} is equipped with $M=32$ antennas and the target outage is fixed to $\epsilon = 0.1$. Lastly, all the numerical results are averaged over $5\times 10^3$ independent \gls{ue} drops.

%=========================================================================
\subsection{Perfect \gls{csit}}
%=========================================================================

\begin{figure}[t!]
\centering
\begin{subfigure}{0.48\textwidth}
    \centering
    \vspace{-5.2mm} \includegraphics[scale=0.6]{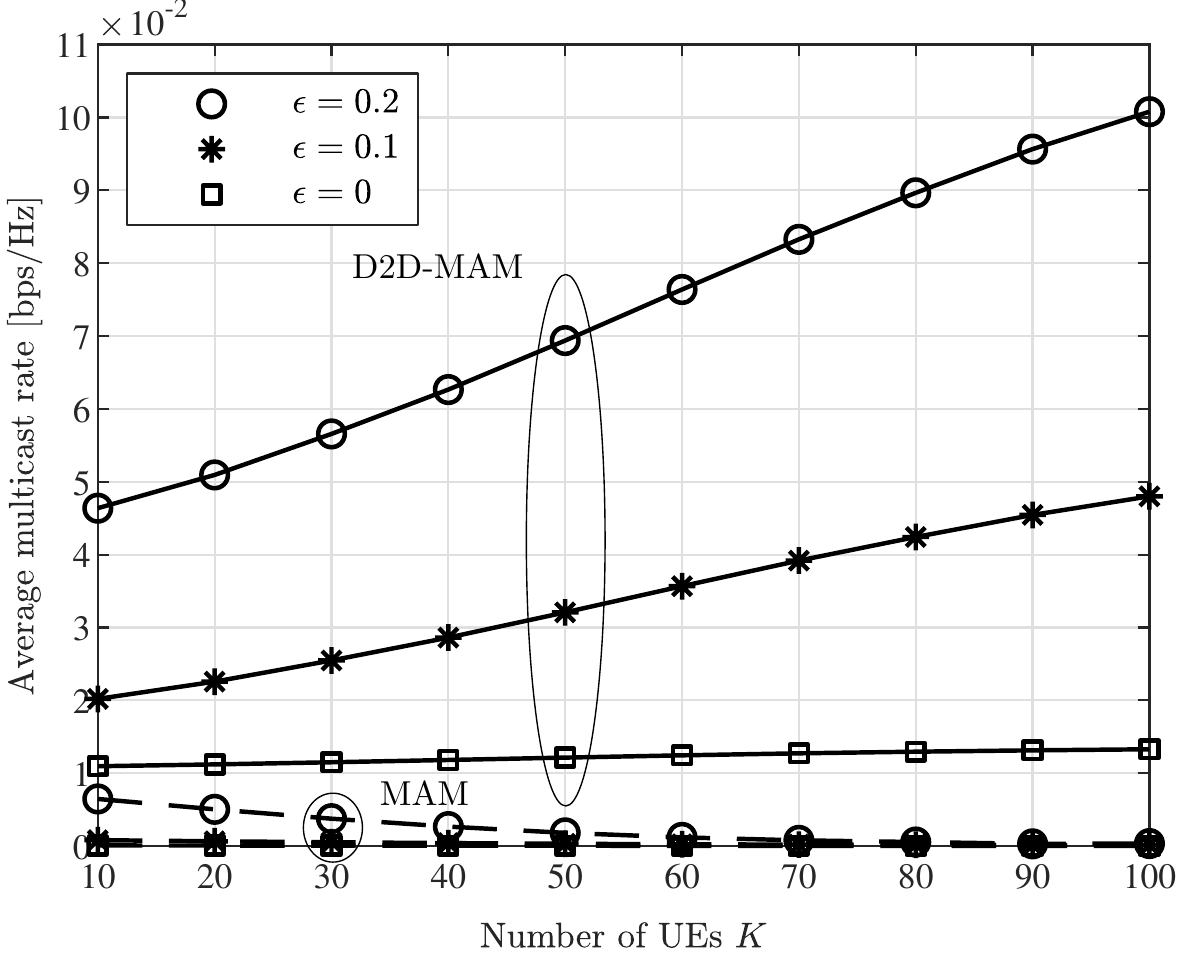}
    \caption{Average multicast rate against the number of \glspl{ue} with $M=32$ and for different values of $\epsilon$.} \label{fig:ICC_r_vs_K_eps}
\end{subfigure} \hfill
\begin{subfigure}{0.48\textwidth}
    \centering
    \includegraphics[scale=0.772]{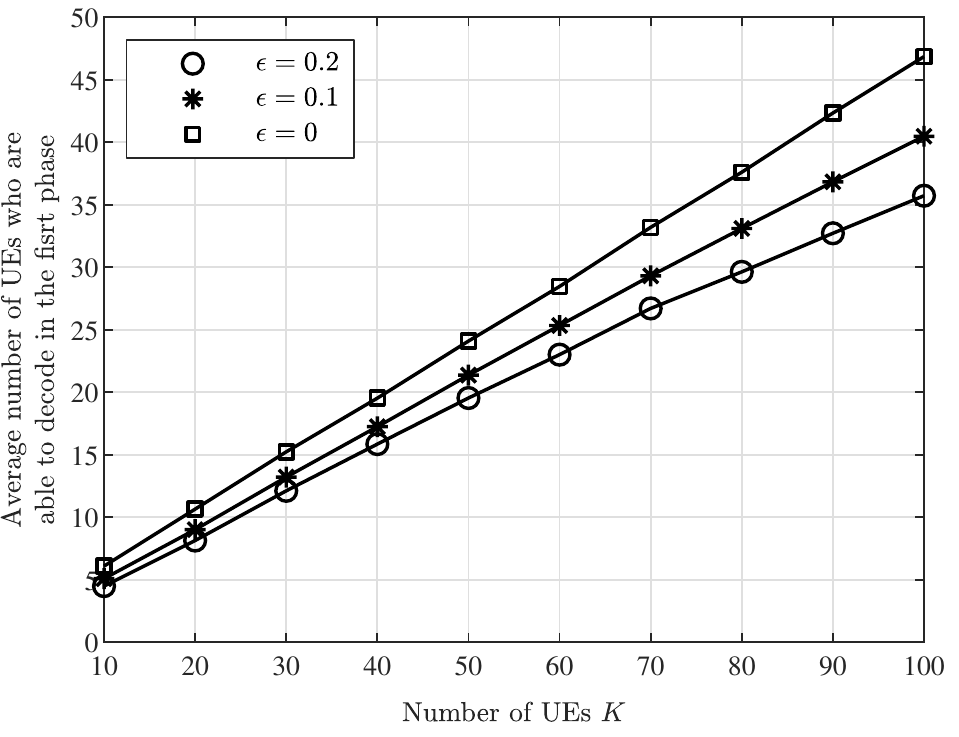}
    \caption{Average number of \glspl{ue} who are able to decode in the first phase against the number of \glspl{ue} with $M=32$ and for different values of $\epsilon$.} \label{fig:ICC_N1_vs_K_eps}
\end{subfigure}
\begin{subfigure}{0.48\textwidth}
    \centering
    \includegraphics[scale=0.6]{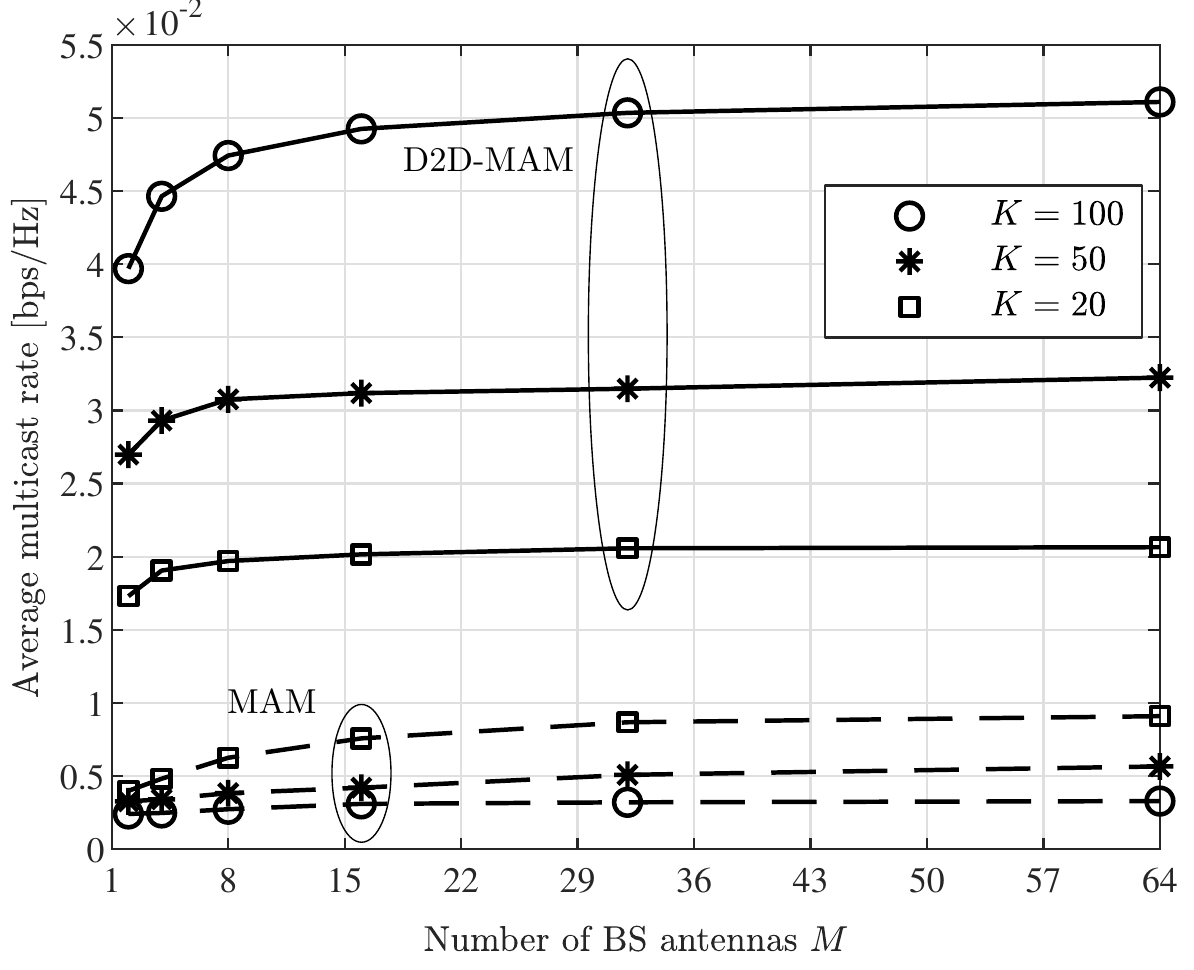}
    \caption{Average multicast rate against the number of \gls{bs} antennas with $\epsilon = 0.1$ and for different values of $K$.} \label{fig:ICC_r_vs_M_K}
\end{subfigure}
\caption{Perfect \gls{csit}: \nameP{} algorithm versus MAM algorithm.} \label{fig:perfect_CSIT}
\end{figure}

In the case of perfect \gls{csit}, we evaluate the performance of the proposed \nameP{} algorithm in Algorithm~\ref{alg:A1} versus the single-phase MAM algorithm described in Section~\ref{subsec:D2D-MAM_bl}. Interestingly, the \nameP{} algorithm converges in very few iterations (typically between $3$ and $10$) even for large values of $K$. Fig.~\ref{fig:perfect_CSIT}(a) plots the average multicast rate against the number of \glspl{ue} for different values of $\epsilon$. Indeed, the second phase of \gls{d2d} communications brings substantial gains with respect to traditional multi-antenna multicasting. In particular, the average multicast rate obtained with the \nameP{} algorithm increases with $K$, whereas that resulting from the MAM algorithm quickly vanishes. Hence, the \nameP{} algorithm effectively overcomes the worst-\gls{ue} bottleneck behavior of conventional single-phase multicasting and remarkably achieves an increasing trend of the multicast rate. In the same setting of Fig.~\ref{fig:perfect_CSIT}(a), Fig.~\ref{fig:perfect_CSIT}(b) shows that the average number of \glspl{ue} who are able to decode in the first phase varies between $35\%$ and $50\%$ of the total \glspl{ue} depending on the target outage. Lastly, Fig.~\ref{fig:perfect_CSIT}(c) illustrates the average multicast rate against the number of \gls{bs} antennas for different values of $K$. Evidently, the \gls{bs} can better focus its transmit power as $M$ increases, which results in an overall improved performance. Here, the lowest value corresponds to $M=1$, i.e., when the \gls{bs} has no beamforming capability and can only transmit in an isotropic fashion in the first phase.

%=========================================================================
\subsection{Statistical \gls{csit}}
%=========================================================================

\begin{figure}[t!]
\centering
\begin{subfigure}{0.48\textwidth}
    \centering
    \includegraphics[scale=0.61]{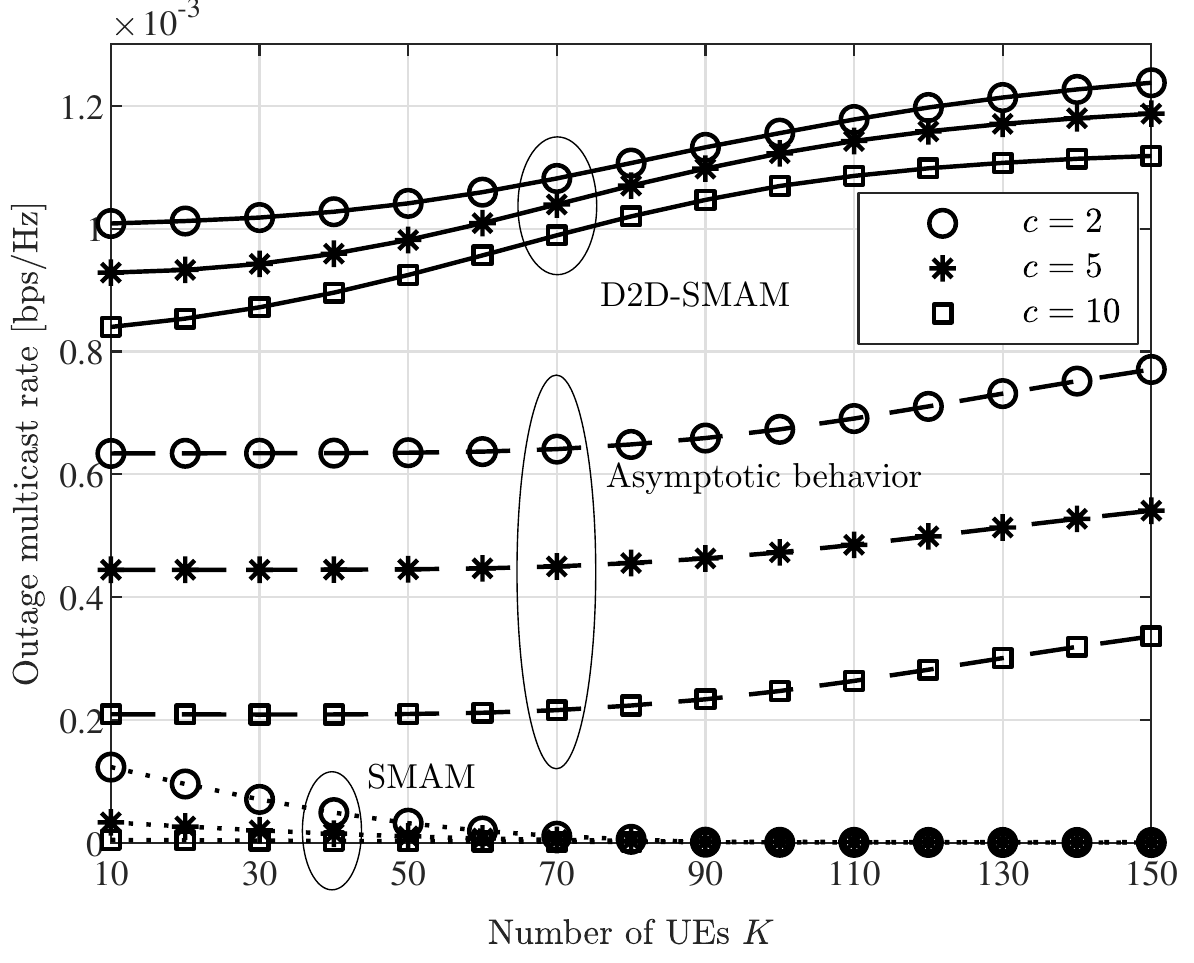}
    \caption{Outage multicast rate against the number of \glspl{ue} for different values of $c = \frac{K}{M}$.} \label{fig:ASI_r_vs_KM}
\end{subfigure} \hfill
\begin{subfigure}{0.48\textwidth}
    \centering
    \includegraphics[scale=0.61]{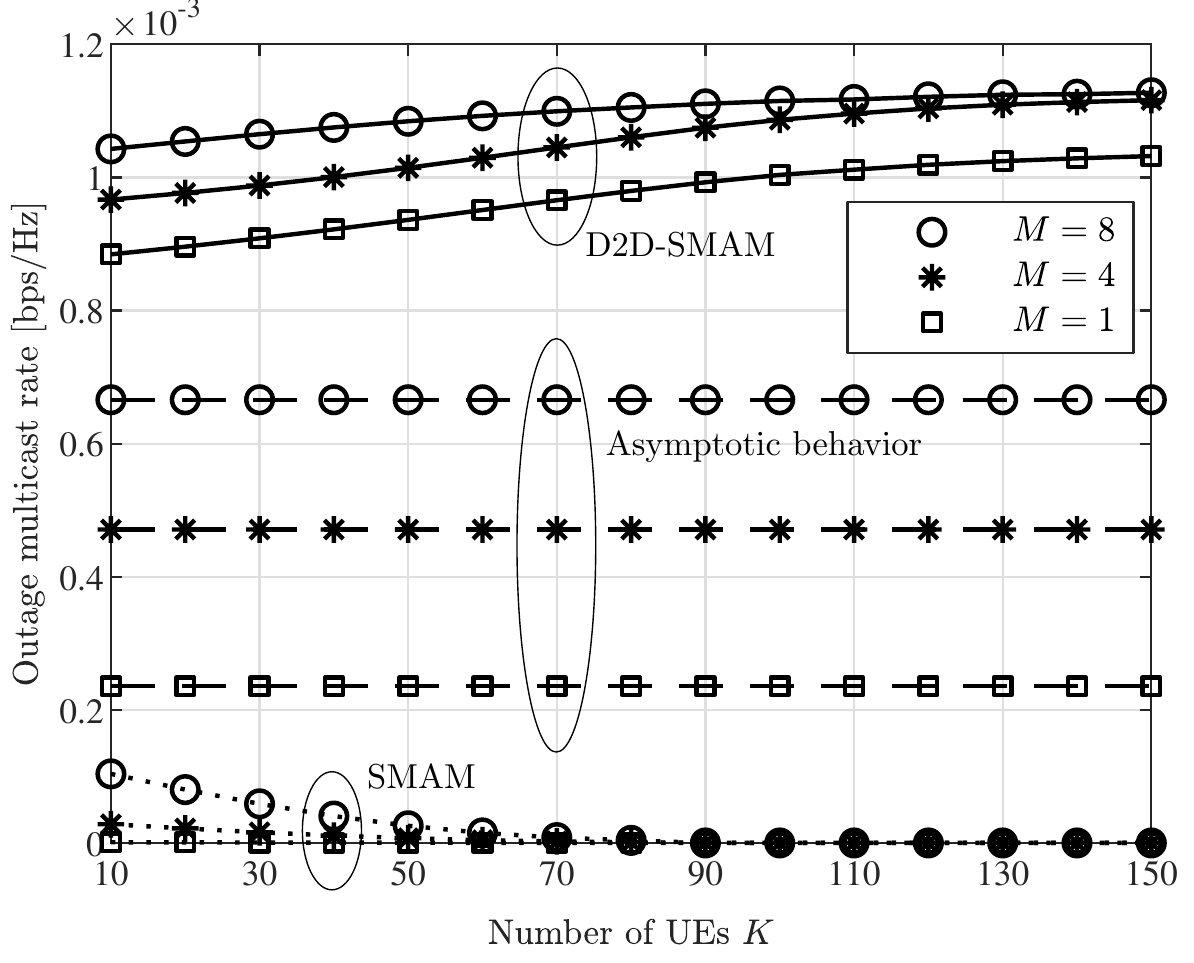}
    \caption{Outage multicast rate against number of \glspl{ue} for different values $M$.} \label{fig:ASI_r_vs_K_M}
\end{subfigure}
\begin{subfigure}{0.48\textwidth}
    \centering
    \includegraphics[scale=0.61]{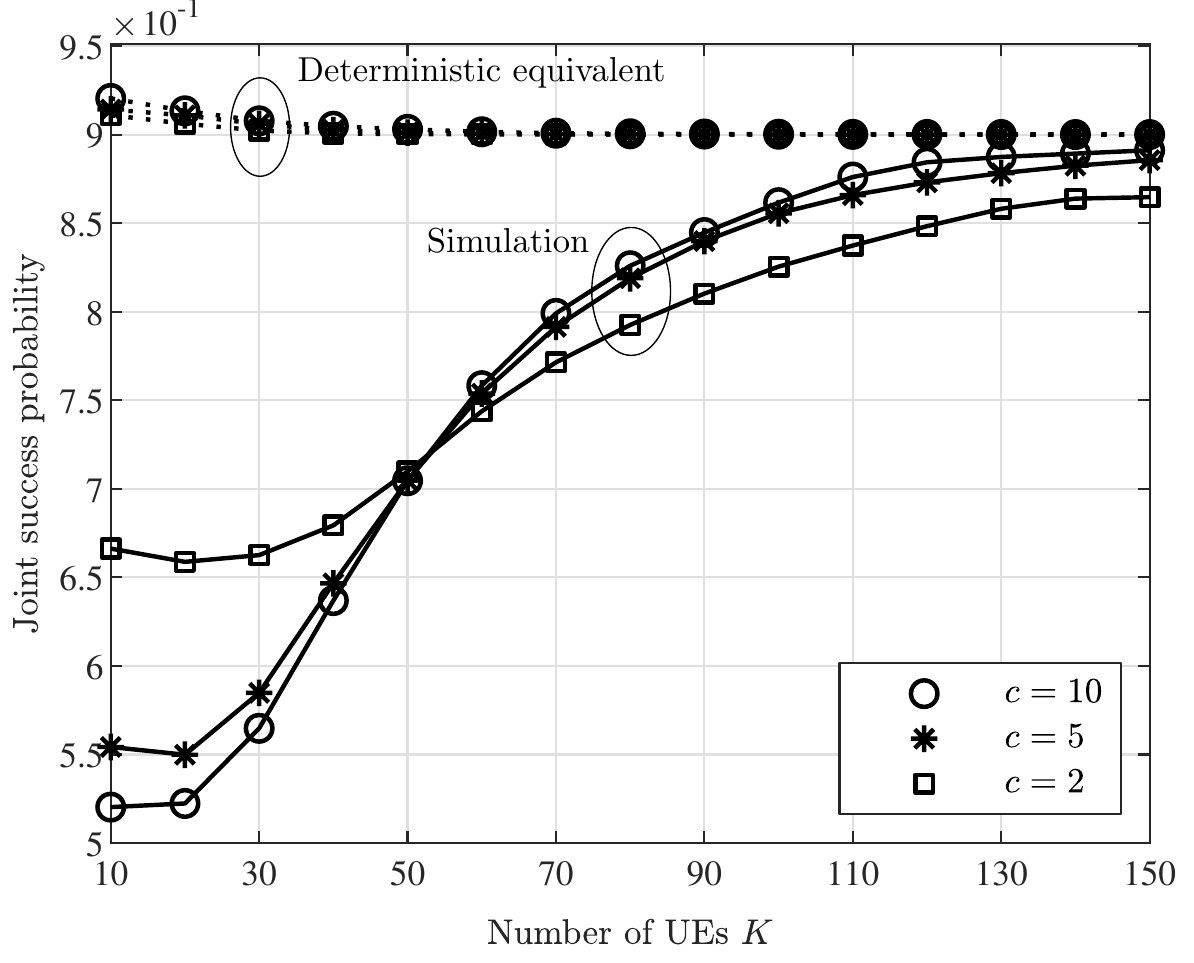}
    \caption{Joint success probability in \eqref{eq:Prob_ph2J} and its deterministic equivalent in \eqref{eq:pjoint2_K} for different values of $c = \frac{K}{M}$.} \label{fig:ASI_Pout_vs_KM}
\end{subfigure}
\caption{Statistical \gls{csit}: \nameS{} algorithm versus SMAM algorithm.} \label{fig:statistical_CSIT}
\end{figure}

In the case of statistical \gls{csit}, we evaluate the performance of the proposed \nameS{} algorithm in Algorithm~\ref{alg:A2} versus the single-phase SMAM described in Section~\ref{sec:D2D-SMAM_bl}. In addition, we compare the asymptotic expressions obtained in Section~\ref{subsec:Asympt} with numerical simulations. For the \nameS{} algorithm, we build the set $\setU$ by identifying $M$ \glspl{ue} whose steering angles satisfy the condition in \eqref{eq:aorth}, while their distance from the \gls{bs} is uniformly distributed. We consider two cases of interest, i.e., where both the number of \glspl{ue} $K$ and the number of \gls{bs} antennas $M$ increase with a fixed ratio $c = \frac{K}{M} >1$ and where $K$ increases for a fixed $M$. The first case is depicted in Fig.~\ref{fig:statistical_CSIT}(a), which shows that the outage multicast rate always grows as long as $M$ grows together with $K$. The second case is illustrated in Fig.~\ref{fig:statistical_CSIT}(b), which shows how increasing $M$ is always beneficial for any given number of \glspl{ue} $K$. Here, the outage multicast rate obtained with the \nameS{} algorithm grows with $K$ and reaches a constant value for large $K$: this is confirmed by its asymptotic behavior, which is constant with $K$. On the contrary, the SMAM algorithm produces a vanishing outage multicast rate and even increasing $M$ does not fundamentally solve this issue. Lastly, Fig.~\ref{fig:statistical_CSIT}(c) compares the joint success probability in \eqref{eq:Prob_ph2J} with its deterministic equivalent in \eqref{eq:pjoint2_K} for different values of $c$. Here, the approximation is tight for sufficiently large values of $K$.

%=========================================================================
\subsection{Topological \gls{csit}}
%=========================================================================

\begin{figure}[t!]
\centering
\begin{subfigure}{0.48\textwidth}
    \centering
    \vspace{2mm}
    \includegraphics[scale=0.45]{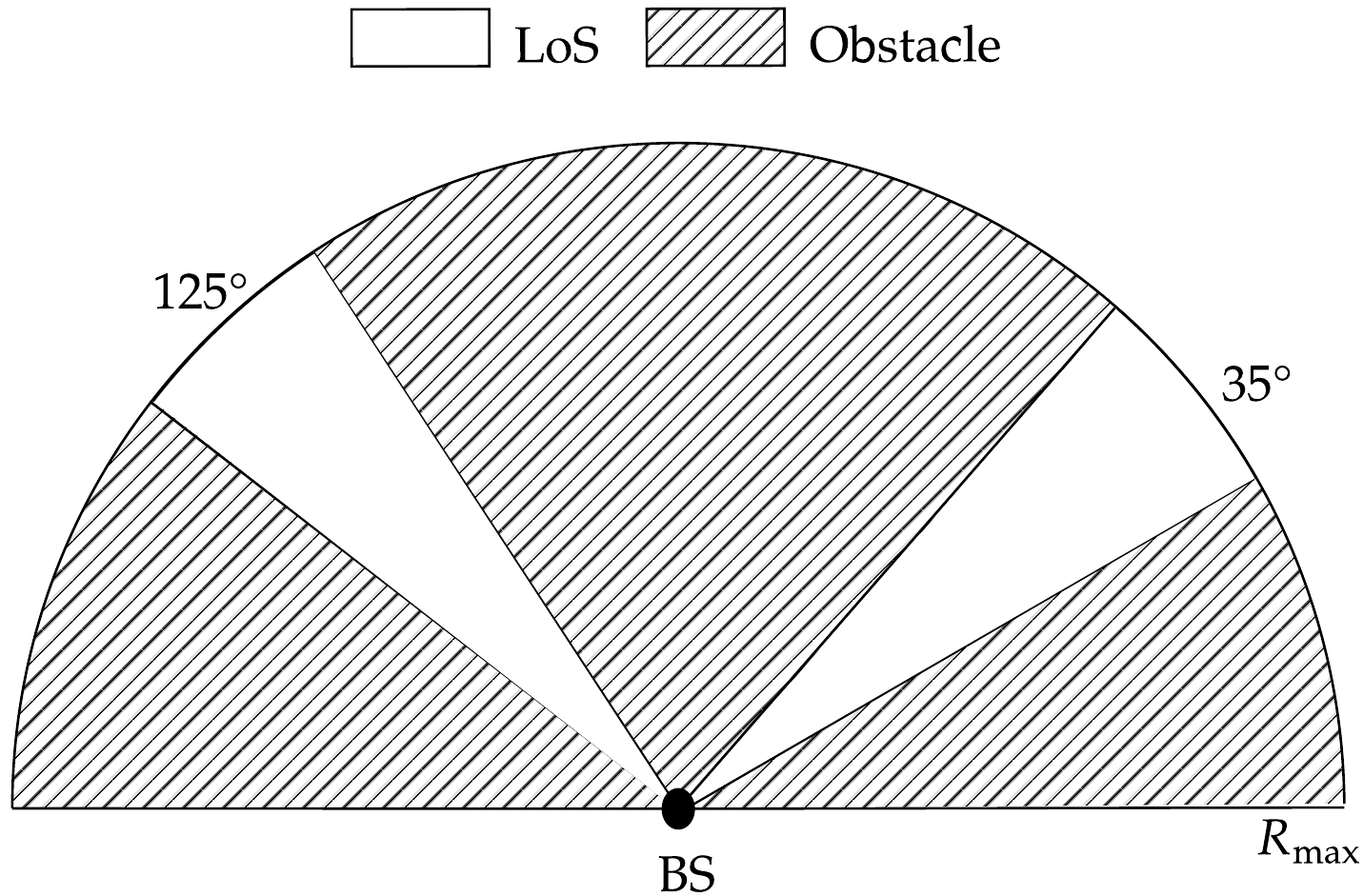}
    \caption{Evaluation scenario.} \label{fig:scenario_toy}
\end{subfigure} \hfill
\begin{subfigure}{0.48\textwidth}
    \centering
    \includegraphics[scale=0.86]{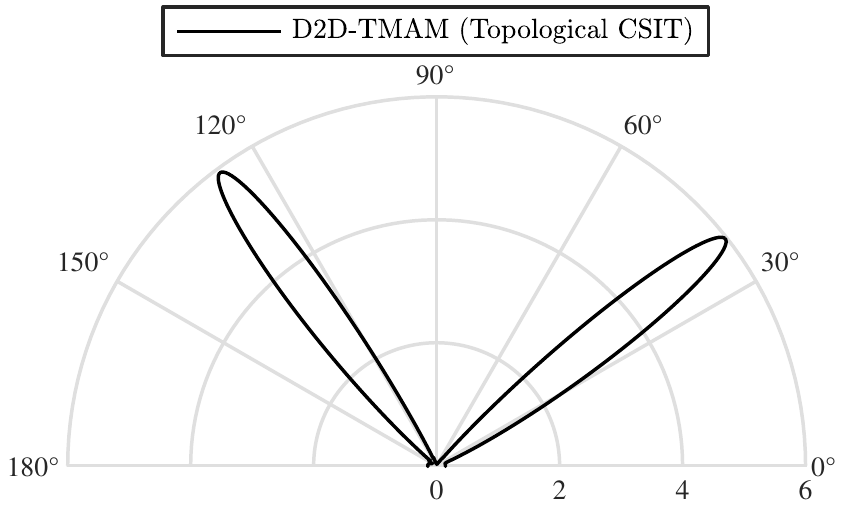}
    \caption{Antenna diagram of the transmit covariance with $M=32$.} \label{fig:Transmit_covar_toy}
\end{subfigure}
\caption{Toy example with topological \gls{csit}: the \glspl{ue} are admitted only in the two white sectors.} \label{fig:toy1}
\end{figure}

In the case of topological \gls{csit}, we evaluate the performance of the proposed \nameT{} algorithm in Algorithm~\ref{alg:A3} versus the \nameP{} algorithm in Algorithm~\ref{alg:A1}, where the latter is based on the assumption of perfect \gls{csit}. Although unfair to the \nameT{} algorithm, this comparison demonstrates how the proposed approach with topological \gls{csit} can accurately sample the long-term network statistics. In turn, this enables to effectively design the precoding strategy at the \gls{bs} with minimal \gls{csit} requirements and no training overhead without excessively compromising the performance. Let $A$ denote the area of the network excluding the regions occupied by the buildings (expressed in m$^2$) and let us consider a uniform \gls{ue} distribution with density $\lambda$ (expressed in \glspl{ue}/m$^2$). In this setting, we assume that each \gls{ue} drop consists of $K$ \glspl{ue}, where $K$ is a Poisson random variable with mean $\bar{K} = \lambda A$. Recall that, for the \nameT{} algorithm, the transmit covariance and the multicast rate are computed offline by averaging the output of the \nameP{} algorithm over $L$ batches of $T$ test points, where we fix $L = 10^3$; on the other hand, the \nameP{} algorithm is executed for each \gls{ue} drop. \vspace{1mm}

\begin{figure}[t!]
\centering
\begin{subfigure}{0.48\textwidth}
    \centering
    \vspace{1.2mm} \includegraphics[scale=0.775]{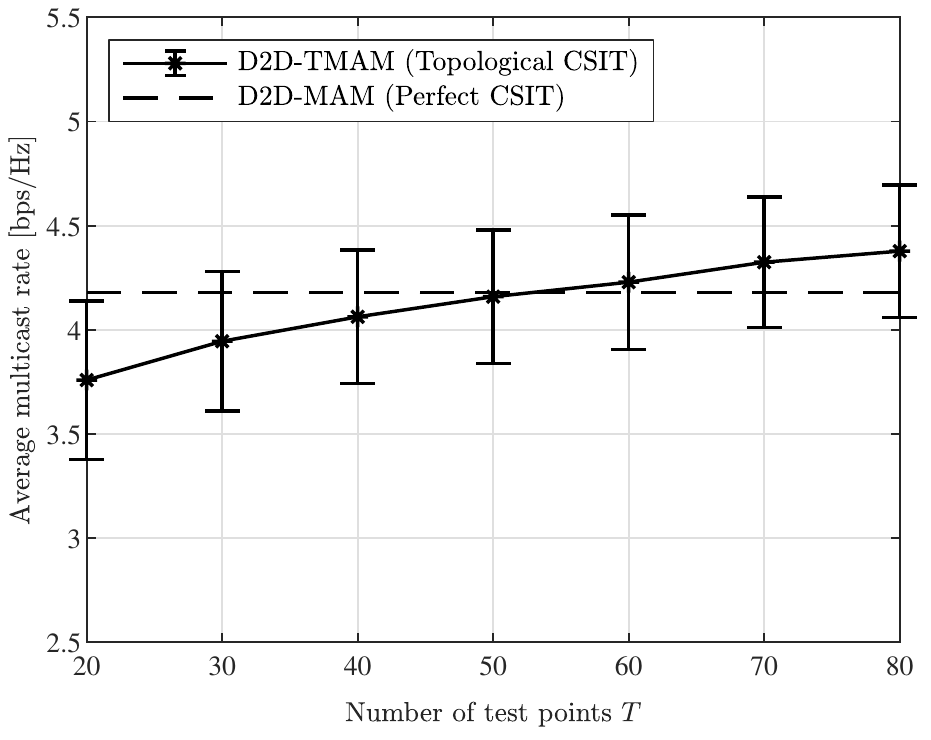}
    \caption{Average multicast rate against number of test points with $M=32$.} \label{fig:multicast_rate_toy}
\end{subfigure} \hfill
\begin{subfigure}{0.48\textwidth}
    \centering
    \includegraphics[scale=0.775]{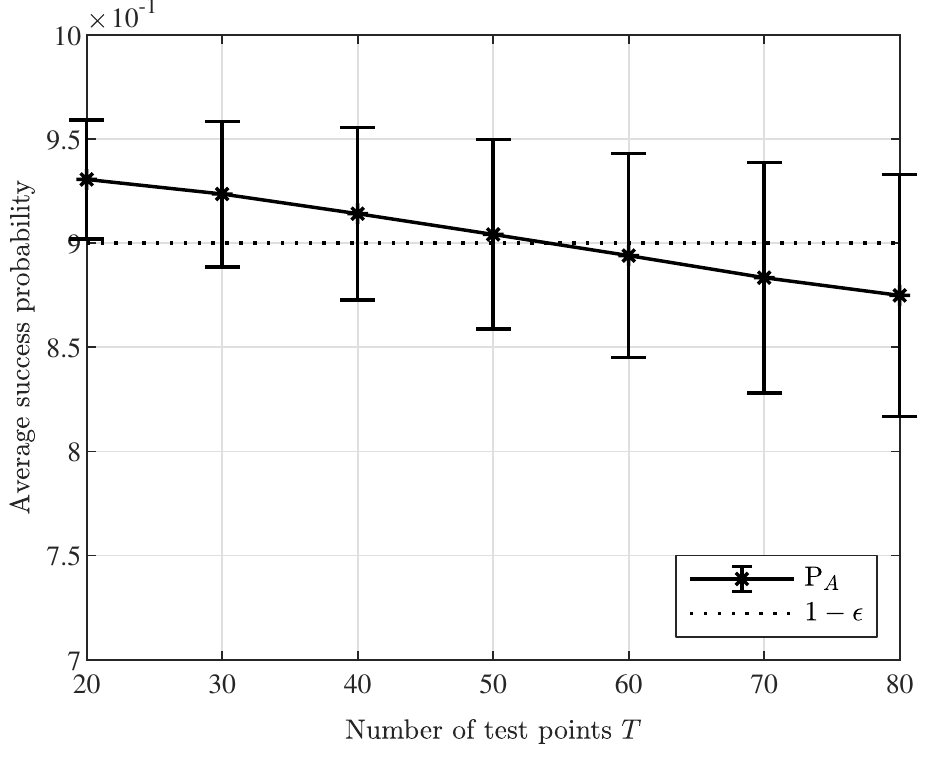}
    \caption{Average success probability against number of test points with $M=32$.} \label{fig:Av_succ_prob_toy}
\end{subfigure}
\caption{Topological \gls{csit} applied to the toy example in Fig.~\ref{fig:toy1}(a): \nameT{} algorithm versus \nameP{} algorithm, where the latter relies on perfect \gls{csit}.} \label{fig:toy2}
\end{figure}

\begin{itemize}
    \item[$\bullet$] \textbf{Toy example.} As a first experiment to verify the effectiveness of the proposed method, we consider the simplified network topology depicted in Fig.~\ref{fig:toy1}(a), with $R_{\mathrm{max}} = 20$~m and where only two sectors admit the presence of \glspl{ue}. In this setting, we have $A = 100$~m$^2$ and, fixing $\lambda = 0.5$~\glspl{ue}/m$^2$, the average number of \glspl{ue} in the network is $\bar{K} =50$; moreover, we assume that all the links are in \gls{los} conditions. Fig.~\ref{fig:toy1}(b) shows the antenna diagram of the transmit covariance obtained with the \nameT{} algorithm with $T = \bar{K}$ test points for each batch: as expected, the multi-antenna beam pattern uniformly covers the two sectors in which the \glspl{ue} are concentrated. Now, we evaluate the average multicast rate and the average success probability as $T$ varies in order to verify which value gives the best performance. Fig.~\ref{fig:toy2} shows that, when $T$ is too small, the algorithm is overcautious and selects a low multicast rate corresponding to an average success probability above the target; on the other hand, when $T$ is too large, the algorithm is overaggressive and selects a high multicast rate corresponding to an average success probability below the target. As expected, the target outage is reached for $T = \bar{K}$ and the corresponding mean value of the average multicast rate is very close to that obtained with the \nameP{} algorithm (which relies on perfect \gls{csit}).
\end{itemize}

\begin{figure}[t!]
\centering
\begin{subfigure}{0.48\textwidth}
    \centering
    \includegraphics[scale=0.615]{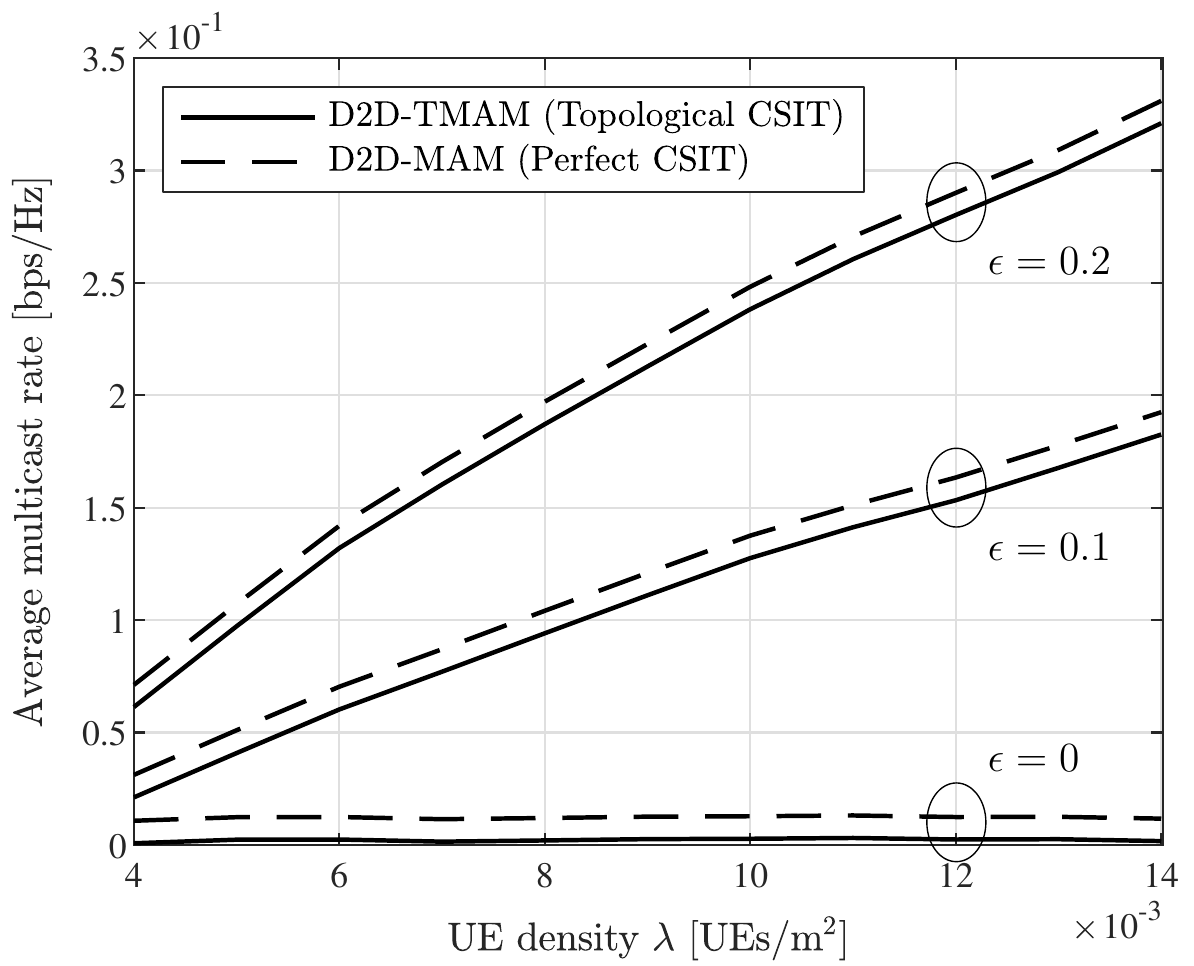}
    \caption{Average multicast rate against \gls{ue} density with $M=32$ and for different values of $\epsilon$.} \label{fig:multicast_rate}
\end{subfigure}
\hfill
\begin{subfigure}{0.48\textwidth}
    \centering
    \includegraphics[scale=0.615]{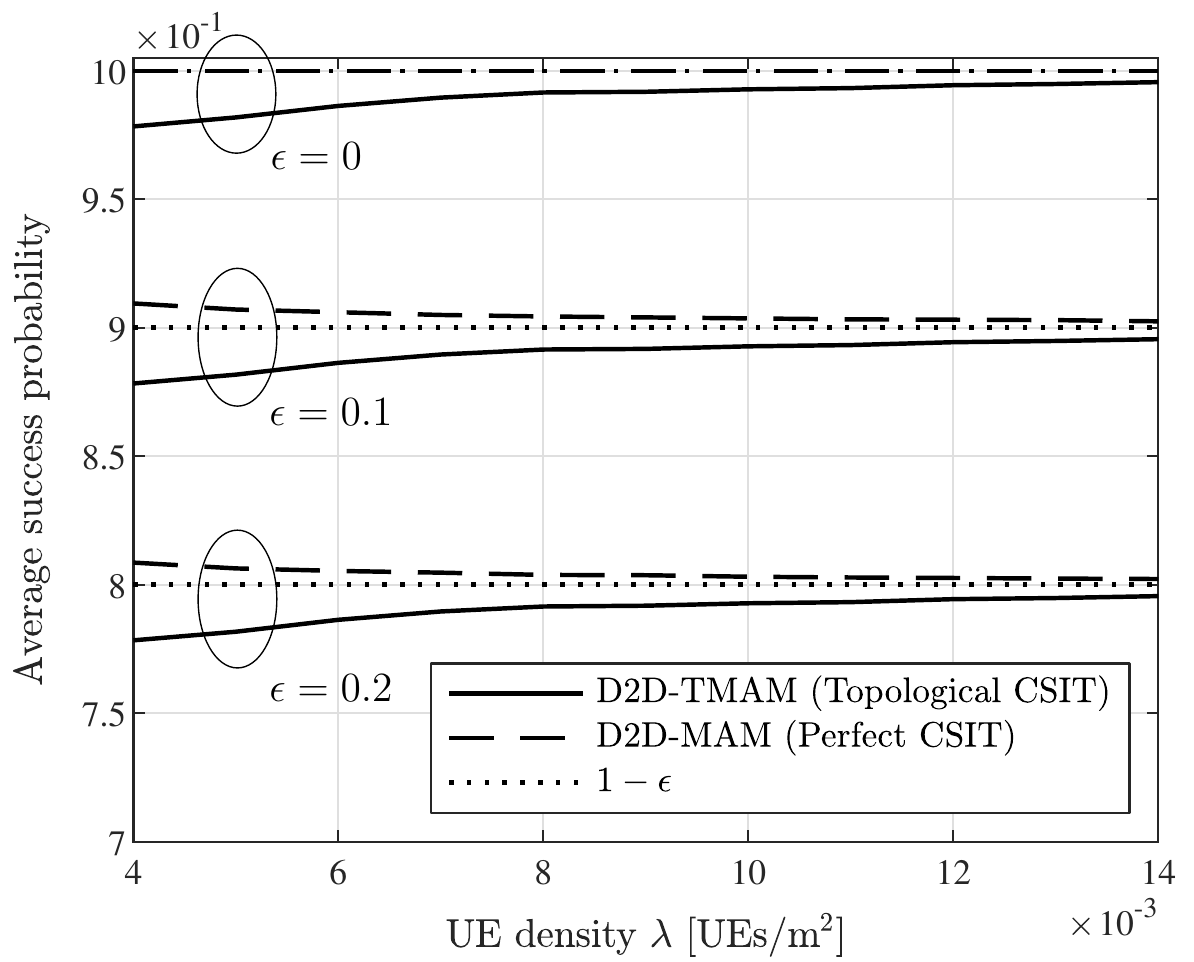}
    \caption{Average success probability against \gls{ue} density with $M=32$ and for different values of $\epsilon$.}
    \label{fig:Av_succ_prob}
\end{subfigure}
\caption{Topological \gls{csit} applied to the evaluation scenario in Fig.~\ref{fig:scenario}: \nameT{} algorithm versus \nameP{} algorithm, where the latter relies on perfect \gls{csit}.} \label{fig:topological_CSIT}
\end{figure}

Now, let us go back to the original evaluation scenario depicted in Fig.~\ref{fig:scenario} and compare the proposed \nameT{} algorithm with the \nameP{} algorithm. Fig.~\ref{fig:topological_CSIT}(a) illustrates the average multicast rate against the \gls{ue} density for different values of $\epsilon$. First of all, we observe that both schemes benefit from increasing the number of \glspl{ue}, thus effectively overcoming the worst-\gls{ue} bottleneck behavior of conventional single-phase multicasting. Furthermore, the performance gap between the \nameT{} algorithm and the \nameP{} algorithm is remarkably small despite the huge difference in the \gls{csit} requirements of the two schemes. Lastly, Fig.~\ref{fig:topological_CSIT}(b) plots the average success probability against the \gls{ue} density for different values of $\epsilon$, showing that the target success probability is achieved more accurately by the \nameT{} algorithm as the \gls{ue} density increases.

%=========================================================================
\section{Conclusion}\label{sec:con}
%=========================================================================

This paper proposes a general two-phase cooperative multicasting framework that leverages both multi-antenna transmission at the \gls{bs} and \gls{d2d} communications between the \glspl{ue}. We explicitly optimize the precoding strategy at the \gls{bs} and the multicast rate over the two phases subject to some outage constraint. In particular, we devise efficient algorithms to tackle three different \gls{csit} configurations, i.e., perfect \gls{csit}, statistical \gls{csit}, and topological \gls{csit}. Numerical results show that the proposed schemes significantly outperform conventional single-phase multi-antenna multicasting in all the considered \gls{csit} configurations. Remarkably, they allow to effectively overcome the vanishing behavior of the multicast rate and achieve an increasing performance as the \gls{ue} population grows large.

\appendices

%=========================================================================
\section{Proof of Proposition~\ref{prop:aorth}}\label{ap:Gamma_orth}
%=========================================================================

Since problem~\eqref{eq:p1_bl} is convex, a given $\Gammab_1$ is optimal if and only if it satisfies the Karush–Kuhn– \linebreak Tucker (KKT) conditions. Let us define the Lagrangian and its gradient as
\begin{align}
    \Lc(\Gammab_1, \mu, \Psib) & \triangleq \sum_{k \in \setK} \frac{1}{\gamma_k \a_k^{\herm} \Gammab_1 \a_k} + \mu \big( \tr(\Gammab_1) - 1 \big) - \tr(\Psib\Gammab_1), \\
    \nabla \Lc(\Gammab_1, \mu, \Psib) & \triangleq - \sum_{k \in \setK} \frac{1}{\gamma_k ( \a_k^{\herm} \Gammab_1 \a_k)^2 } \a_k \a_k^{\herm} + \mu \I_M - \Psib
\end{align}
respectively, where we have introduced the dual variables $\mu \in \Real$ and $\Psib \in \Compl^{M \times M}$. The KKT conditions of problem~\eqref{eq:p1_bl} can be written as
\begin{subequations} \label{eq:440}
\begin{align}
    \displaystyle & \sum_{k \in \setK}  \frac{1}{\gamma_k ( \a_k^{\herm} \Gammab_1 \a_k)^2 } \a_k \a_k^{\herm} = \mu \I_M  - \Psib, \label{eq:440A} \\
    \displaystyle & \tr(\Gammab_1) \leq 1, \ \Gammab_1 \succeq \0, \label{eq:440B} \\
    \displaystyle & \mu \geq 0, \ \Psib \succeq \0, \label{eq:440C} \\
    \displaystyle & \mu \big( \tr(\Gammab_1)-1 \big) = 0, \ \Psib\Gammab_1 = \0. \label{eq:440D}
\end{align}
\end{subequations}
The condition in \eqref{eq:440A} suggests that the transmit covariance has the structure
\begin{align}\label{eq:448_prf}
    \Gammab_1 = \sum_{k \in \setK} w_k \a_k \a_k^{\herm} 
\end{align}
where $\sum_{k \in \setK} w_k = 1/M$ implies $\tr(\Gammab_1) =1$ and $\{ w_k \geq 0 \}_{k \in \setK}$ implies $\Gammab_1 \succeq \0$. From \eqref{eq:448_prf}, we can write
\begin{align}
    \a_k^{\herm} \Gammab_1 \a_k & = \sum_{j \in \setK} w_j \phi_{kj} 
\end{align}
where we have defined $\phi_{kj} \triangleq |\a_k^{\herm}\a_j|^2$, with $\Phib \triangleq [\phi_{kj}]_{k,j \in \setK} \in \Compl^{K \times K}$ being a symmetric matrix with diagonal elements equal to $M^2$. Plugging \eqref{eq:448_prf} into \eqref{eq:440}, the KKT conditions become
\begin{subequations} \label{eq:KKT2}
\begin{align}
    \displaystyle & \sum_{k \in \setK} \frac{1}{\gamma_k \big( \sum_{j \in \setK} w_j \phi_{kj} \big)^2} \a_k \a_k^{\herm} = \mu \I_M - \Psib, \label{eq:KKT2A} \\
    \displaystyle & \sum_{k \in \setK} w_k = \frac{1}{M}, \ \{ w_k \geq 0 \}_{k \in \setK}, \label{eq:KKT2B} \\
    \displaystyle & \mu \geq 0, \ \Psib \succeq \0, \label{eq:KKT2C} \\
    \displaystyle & \mu \bigg( \sum_{k \in \setK} w_k - \frac{1}{M} \bigg) = 0, \ \Psib \sum_k w_k \a_k \a_k^{\herm} = \0. \label{eq:KKT2D}
\end{align}
\end{subequations}
Let us define $\w \triangleq [w_1,\ldots,w_K]^{\tran} \in \Real^{K\times 1}$. Choosing the weights that satisfy \eqref{eq:KKT2B} allows us to set $\Psib = \0$ and, from \eqref{eq:KKT2A}, we can show that
\begin{align} \label{eq:w_opt_prf}
    \w = \frac{1}{\sqrt{\mu M}} \Phib^{-1} \b
\end{align}
where we have defined 
\begin{align}
    \b \triangleq \bigg[ \frac{1}{\sqrt{\gamma_1 \1^{\tran}\Phib^{-1}\e_1}}, \ldots, \frac{1}{\sqrt{\gamma_K \1^{\tran}\Phib^{-1}\e_K}} \bigg]^{\tran}.
\end{align}
On the other hand, $\mu$ can be obtained by plugging \eqref{eq:w_opt_prf} into the first condition in \eqref{eq:KKT2D}, i.e.,
\begin{align}
    \mu & = M (\1^{\tran} \Phib^{-1} \b)^2 \label{eq:mu}
\end{align}
and, by plugging \eqref{eq:mu} into \eqref{eq:w_opt_prf}, we obtain
\begin{align} \label{eq:w_k}
    w_{k} = \frac{\e_{k}^{\tran} \Phib^{-1} \b}{M \1^{\tran} \Phib^{-1} \b}, \quad \forall k \in \setK.
\end{align}
Finally, choosing $\{ w_{k} \}_{k \in \setK}$ as in \eqref{eq:w_k}, $\mu$ as in \eqref{eq:mu}, and $\Psib = \0$ readily satisfies \eqref{eq:KKT2B}--\eqref{eq:KKT2D}, whereas \eqref{eq:KKT2A} yields
\begin{align}\label{eq:KKT_verify2}
    \sum_{k \in \setK} (\1^{\tran} \Phib^{-1} \e_{k}) \a_k \a_k^{\herm} = \frac{1}{M} \I_M.
\end{align}
The latter is satisfied when $\Phib = M^2 \I_K$, i.e., when $K = M$ and the steering angles of the \glspl{ue} are such that $\a_k^{\herm} \a_j = 0$, $\forall k \neq j$ (see, e.g., \cite{Saye02} for more details). In this setting, it follows from \eqref{eq:w_k} that $w_{k} = 1/(M \sqrt{\gamma_{k}} \nu_{\setK})$, from which we obtain the expression of the optimal transmit covariance in \eqref{eq:gamma_aorth}. \hfill $\IEEEQED$

\addcontentsline{toc}{chapter}{References}
\bibliographystyle{IEEEtran}
\bibliography{IEEEabrv,refs}

\end{document}

%% file: Definitions.tex
\renewcommand{\a}{\mathbf{a}}
\renewcommand{\b}{\mathbf{b}}

\newcommand{\e}{\mathbf{e}}

\newcommand{\h}{\mathbf{h}}

\newcommand{\p}{\mathbf{p}}

\newcommand{\w}{\mathbf{w}}
\newcommand{\x}{\mathbf{x}}

\newcommand{\z}{\mathbf{z}}

\newcommand{\0}{\mathbf{0}}
\newcommand{\1}{\mathbf{1}}

\renewcommand{\H}{\mathbf{H}}
\newcommand{\I}{\mathbf{I}}

\newcommand{\Gammab}{\mathbf{\Gamma}}

\newcommand{\Phib}{\mathbf{\Phi}}
\newcommand{\Psib}{\mathbf{\Psi}}

\newcommand{\setA}{\mathcal{A}}

\newcommand{\setC}{\mathcal{C}}

\newcommand{\setK}{\mathcal{K}}

\newcommand{\setN}{\mathcal{N}}

\newcommand{\setU}{\mathcal{U}}

\newcommand{\Real}{\mbox{$\mathbb{R}$}}
\newcommand{\Compl}{\mbox{$\mathbb{C}$}}

\newcommand{\argmax}{\operatornamewithlimits{argmax}}

\newcommand{\diff}{\mathrm{d}}
\newcommand{\Exp}{\mathbb{E}}

\newcommand{\herm}{\mathrm{H}}

\renewcommand{\Pr}{\mathbb{P}}

\newcommand{\tr}{\mathrm{tr}}
\newcommand{\tran}{\mathrm{T}}